\documentclass{shep3}

\usepackage{axodraw,epsfig}
\usepackage{cite}
  \usepackage{graphicx}
  \usepackage{amssymb}

\preprint{
  ZU-TH 16/08\\
}

\title{Two-Loop QCD Corrections to the
Heavy-to-Light Quark Decay}

\author{R.~Bonciani$\rm \, ^{a,}$\footnote{Email: {\tt
Roberto.Bonciani@physik.uzh.ch}} \, and  A.~Ferroglia$\rm \, ^{a,
}$\footnote{Email: {\tt
Andrea.Ferroglia@physik.uzh.ch}} \\

{\it $\rm ^a$ Institut f{\"u}r
Theoretische Physik,
Universit{\"a}t Z\"urich,
CH-8057 Zurich, Switzerland}\\

}

\abstract{We present an analytic expression for the two-loop QCD
corrections to the decay process  $b \to u \, W^{*} $, where $b$
and $u$ are a massive and massless quark, respectively, while
$W^{*}$ is an off-shell charged weak boson. Since the $W$-boson
can subsequently decay in a lepton  anti-neutrino pair, the
results of this paper are a first step towards a fully analytic
computation of differential distributions for the semileptonic
decay of a $b$-quark. The latter partonic process plays a crucial
role in the study of inclusive semileptonic charmless decays of
$B$-mesons. The three independent form factors characterizing the
$b W u$ vertex are provided in form of a Laurent series in
$(d-4)$, where $d$ is the space-time dimension. The coefficients
in the series are expressed in terms of Harmonic Polylogarithms of
maximal weight 4, and are functions of the invariant mass of the
leptonic decay products of the $W$-boson.}

\keywords{Heavy Quark Decay, Two-Loop Calculations}

\begin{document}

\newcommand{\bfig}{\begin{figure}}
\newcommand{\efig}{\end{figure}}
\newcommand{\be}{\begin{equation}}
\newcommand{\ee}{\end{equation}}
\newcommand{\bea}{\begin{eqnarray}}
\newcommand{\eea}{\end{eqnarray}}
\newcommand{\bc}{\begin{center}}
\newcommand{\ec}{\end{center}}
\newcommand{\nn}{\nonumber}

\section{Introduction}

The measurements of  inclusive semileptonic $B$ meson decays, such as $\bar{B} \to X_u
\, l \, \bar{\nu}$ and  $\bar{B} \to X_c \, l  \,\bar{\nu}$, allow a precise
determination of the CKM matrix elements $|V_{ub}|$ and $|V_{cb}|$. The latter
are relevant for  the study of flavor and CP violation in the quark sector (for
a recent review see \cite{Gardi:2008bb}).

Total decay rates of the $B$ meson are described by a local Operator Product
Expansion (OPE) in inverse powers of the $b$-quark mass $m_b$.  To leading order
in $1/m_b$, the total $B$ meson decay rate is equivalent  to the decay rate of
an on-shell $b$ quark, which can be calculated in perturbation theory
\cite{HQE}. Many authors contributed to the calculation of the radiative
corrections to the total decay rate of $b \to u \, l \, \bar{\nu}$ and  $b
\to c \, l \, \bar{\nu}$, at ${\mathcal O}(\alpha_S)$
\cite{bu1loop,bc1loop} and ${\mathcal O}(\alpha_S^2)$
\cite{Luke:1994du,Ball:1995wa,Czarnecki:1997cf,Franzkowski:1997vg,Blokland:2004ye,
Pak:2006xf,vanRitbergen:1999gs,Steinhauser:1999bx}.  However, experimental 
collaborations need to
impose cuts (also severe) on the kinematic variables. For instance, in charmless
semileptonic decays, the need to suppress the charm background (which is $\sim
50$ times larger than the signal) forces one to restrict the measurements to the
``shape-function region'', in which the hadronic final state has large energy
($E_X \sim m_b$), but only moderate invariant mass ($ \sim m_b \Lambda_{QCD}$).
It is therefore of great interest to consider differential decay distributions,
from which it is possible to derive predictions for partial decay rates with
arbitrary cuts. In this context, a first important set of results  was obtained
in \cite{DFN}, where it is possible to find analytic expressions for the NLO
triple-differential distribution of the semileptonic $\bar{B} \to X_u \, l \,
\bar{\nu}$ decay together with several double and single differential
distributions for the same process.  The resummation of threshold logarithms  to
next-to-leading approximation  in the $b \to u$ transition was considered in
\cite{Aglietti:2001br}.  Higher order contributions to $\bar{B} \to X_u \, l \,
\bar{\nu}$ decays were considered in
\cite{Gardi:2005yi,Andersen:2005mj,Aglietti:2005mb,Gambino:2006wk} and, very
recently, the full NNLO QCD corrections to the  partonic process $b \to c \,  l
\, \bar{\nu}$ were obtained in \cite{Melnikov:2008qs}.
Since the OPE applies only to sufficiently inclusive quantities, different
frameworks were developed in order to account for effects due to cuts on the
kinematic space \cite{Bauer:2000yr,Bauer:2001yt,Bosch:2004th,Lange:2005yw,
Ligeti:2008ac,Aglietti:2007bp,Gambino:2007rp}.
In particular, in the shape-function region, Soft Collinear Effective Theory
(SCET)  provides an appropriate framework for the evaluation of the
triple-differential distribution of the inclusive semileptonic decay $\bar{B} \to X_u
\, l \, \bar{\nu}$. The NLO analysis of the latter process  within the SCET
approach is presented in \cite{Bosch:2004th,Lange:2005yw}.
At NNLO, the situation is more complicated, but the jet and soft functions  are
known to ${\mathcal O}(\alpha_S^2)$ in perturbation theory
\cite{Becher:2005pd,Becher:2006qw}. The only missing piece is  the hard
function, which can be obtained from the two-loop QCD corrections to the decay
of a $b$-quark into a $u$-quark and an off-shell $W$-boson \cite{Pecjak}.  On
the other hand, these virtual corrections can be considered as a first step
towards an exact evaluation of the  NNLO QCD corrections to the heavy-to-light
quark transition. To complete the latter calculation, it is also necessary to
take into account the real emission.

In this work we focus on the calculation of the two-loop QCD corrections to the
decay process $b \to u \, W^{*} $. We provide an analytic expression for the
three independent vertex form factors characterizing the coupling of the quark
current with the charged weak boson. These form factors are evaluated by
employing a set of techniques which are by now standard in multiloop
calculations (see for instance \cite{ffact}). We generate the relevant Feynman
diagrams with QGRAF \cite{QGRAF}. The form factors are extracted directly from
the Feynman diagrams by means of projector operators. The whole calculation is
carried  out in Dimensional Regularization (DR); UV and IR (soft  and
collinear)  divergencies appear as poles in $(d-4)$, where $d$ is the space-time
dimension. Since we work in DR, a prescription for handling the matrix
$\gamma_5$ in $d$-dimensions must be chosen. We employed a $\gamma_5$ which
anticommutes with $\gamma_{\mu}$ in $d$-dimensions. This prescription is
appropriate for the case under study, since it is known that the diagrams that
we  consider fulfill a canonical (non-anomalous)  Ward identity. After applying
the projectors, the contribution of individual Feynman diagrams to the form
factors is given by a combination of dimensionally regularized scalar integrals.
These integrals  are related  to a small set of master integrals (MIs) by means
of the Laporta algorithm \cite{Laporta}. The MIs are evaluated by employing the
Differential Equations method \cite{DiffEq} and they are expressed as   Laurent
series in $(d-4)$. The coefficients of the series are given in terms of Harmonic
Polylogarithms (HPLs) \cite{HPLs} of a single dimensionless variable
$y=q^2/m_b^2= -M_l^2/m_b^2$, where $q^2$ is the squared momentum carried by the
$W$-boson and $M_l$ is the lepton pair invariant mass.  Since $y$ is negative
in the physical region ($-1 \leq y \leq 0$) we perform an analytic continuation
$y \to - x - i0^+$, where now $x = M_l^2/m_b^2$, $0 \leq x \leq 1$. The analytic
continuation is indeed completely trivial, since all the HPLs appearing in the
result are real for $-1 \leq y \leq 0$.
The form factors found with the above procedure still contain UV and IR
divergencies. It is possible to get rid of the UV divergencies by means of the
renormalization procedure. We renormalize the form factors in a mixed scheme:
the heavy- and light-quark wave functions and the heavy-quark mass are
renormalized in the {\it on-shell} ({\it OS\,}) scheme, while the strong
coupling constant  is renormalized in $\overline{\mbox{MS}}$ scheme. The results
shown in this paper contain IR divergencies. In order to cancel them, it is
necessary to combine these results with the appropriate jet and soft functions
\cite{Bosch:2004th}.  Analogously, one can add the exact real emission and
consider physical observables which are sufficiently inclusive with respect to
the hard and soft radiation.

%

The paper is structured as follows. In section \ref{Fdiag}, we introduce the
Feynman diagrams involved in the calculation and we discuss their structure in
terms of form factors. In section \ref{ren},  it is possible to find the details
of the UV renormalization procedure. In sections \ref{1lFF} and \ref{2lFF}, we
collect the analytic expressions of the UV renormalized one- and two-loop QCD
corrections to the form factors, respectively. The expressions of the bare form 
factors  as well as the contributions of the individual diagrams to the form
factors can be found  in \cite{submission}. In section \ref{Ward}, we discuss
the Ward identity relevant for the $b \to u \, W^* $ decay and  we prove that
our form factors satisfy it. We also provide the analytic expression of the one-
and two-loop QCD corrections to the scalar vertex in which the pseudo-Goldstone
boson couples to the quarks, since it enters in the Ward identity fulfilled by
the $b W u$ vertex. Our conclusions  can be found in section \ref{concl}.
Finally, in appendix \ref{app1} we collect the  set of MIs employed in the
calculation.

\section{Feynman Diagrams and Form Factors \label{Fdiag}}

We consider the decay process $b \to u \, W^{*}  \to u \, l \, \bar{\nu}_l$. The
bottom quark of mass $m_b$ carries a momentum $P$ and decays in an up-quark
(considered as massless) which carries momentum $p$ and a $W$-boson of momentum
$q=P-p$. Subsequently, the $W$-boson decays in the pair $l \bar{\nu}_l$ of
squared invariant mass $M_l^2 = - q^2$.
The mass-shell conditions are such that $P^2 = - m_b^2$ and $p^2 = 0$. The
Feynman diagrams contributing to the two-loop QCD corrections to the decay
process $b \to u \, W^{*} $ are shown in Fig.~\ref{fig1}.
%
\begin{figure}
\vspace*{10mm}
\bc
\[ \vcenter{
\hbox{
  \begin{picture}(0,0)(0,0)
\SetScale{1.}
  \SetWidth{.2}
\Photon(-50,30)(-35,15){2}{4}
\Line(-35,15)(20,15)
\ArrowLine(20,15)(35,15)
\Gluon(-35,-5)(-3,15){-3}{7}
\Gluon(-35,-20)(20,15){-3}{11}
\Text(0,-45)[c]{(a)}
\Text(10,23)[c]{\it p}
\LongArrow(20,23)(35,23)
\Text(-48,-33)[c]{\it P}
\LongArrow(-48,-23)(-48,-8)
\Text(-32,24)[c]{\it q}
\LongArrow(-38,30)(-48,40)
  \SetWidth{1.5}
\ArrowLine(-35,-38)(-35,-23)
\Line(-35,-23)(-35,15)
\end{picture}}
}
\hspace{4.cm}
\vcenter{
\hbox{
  \begin{picture}(0,0)(0,0)
\SetScale{1.}
  \SetWidth{.2}
\Photon(-50,30)(-35,15){2}{4}
\Line(-35,15)(20,15)
\ArrowLine(20,15)(35,15)
\Gluon(-35,-5)(20,15){-3}{10}
\Gluon(-35,-20)(-5,15){-3}{8}
\Text(0,-45)[c]{(b)}
  \SetWidth{1.5}
\ArrowLine(-35,-38)(-35,-23)
\Line(-35,-23)(-35,15)
\end{picture}}
}
\hspace{4.cm}
\vcenter{
\hbox{
  \begin{picture}(0,0)(0,0)
\SetScale{1.}
  \SetWidth{.2}
\Photon(-50,30)(-35,15){2}{4}
\Line(-35,15)(20,15)
\ArrowLine(20,15)(35,15)
\Gluon(-35,-5)(20,15){-3}{10}
\GlueArc(-35,-5)(10,90,270){3}{4}
\Text(0,-45)[c]{(c)}
  \SetWidth{1.5}
\ArrowLine(-35,-38)(-35,-23)
\Line(-35,-23)(-35,15)
\end{picture}}
}
\hspace{4.cm}
\vcenter{
\hbox{
  \begin{picture}(0,0)(0,0)
\SetScale{1.}
  \SetWidth{.2}
\Photon(-50,30)(-35,15){2}{4}
\Line(-35,15)(20,15)
\ArrowLine(20,15)(35,15)
\Gluon(-35,-23)(20,15){-3}{10}
\GlueArc(-35,-5)(10,90,270){3}{4}
\Text(0,-45)[c]{(d)}
  \SetWidth{1.5}
\ArrowLine(-35,-38)(-35,-23)
\Line(-35,-23)(-35,15)
\end{picture}}
}\]
%
%
%
\vspace{2.2cm}
\[\vcenter{
\hbox{
  \begin{picture}(0,0)(0,0)
\SetScale{1.}
  \SetWidth{.2}
\Photon(-50,30)(-35,15){2}{4}
\Line(-35,15)(15,15)
\ArrowLine(15,15)(35,15)
\Gluon(-35,-15)(5,15){-3}{10}
\GlueArc(0,15)(15,0,180){3}{10}
\Text(0,-45)[c]{(e)}
  \SetWidth{1.5}
\ArrowLine(-35,-38)(-35,-23)
\Line(-35,-23)(-35,15)
\end{picture}}
}
%
\hspace{4.cm}
%
\vcenter{
\hbox{
  \begin{picture}(0,0)(0,0)
\SetScale{1.}
  \SetWidth{.2}
\Photon(-50,30)(-35,15){2}{4}
\Line(-35,15)(20,15)
\ArrowLine(20,15)(35,15)
\Gluon(-35,-15)(22,15){-3}{10}
\GlueArc(-10,15)(15,0,180){3}{10}
\Text(0,-45)[c]{(f)}
  \SetWidth{1.5}
\ArrowLine(-35,-38)(-35,-23)
\Line(-35,-23)(-35,15)
\end{picture}}
}
\hspace{4.cm}
%
\vcenter{
\hbox{
  \begin{picture}(0,0)(0,0)
\SetScale{1.}
  \SetWidth{.2}
\Photon(-50,30)(-35,15){2}{4}
\Line(-35,15)(20,15)
\ArrowLine(20,15)(35,15)
\Gluon(-35,-15)(-15,-15){-3}{2}
\Gluon(10,-15)(22,15){-3}{4}
\GlueArc(-2,-15)(13,0,180){3}{4}
\GlueArc(-2,-15)(13,180,360){3}{4}
\Text(0,-45)[c]{(g)}
  \SetWidth{1.5}
\ArrowLine(-35,-38)(-35,-23)
\Line(-35,-23)(-35,15)
\end{picture}}
}
\hspace{4.cm}
\vcenter{
\hbox{
  \begin{picture}(0,0)(0,0)
\SetScale{1.}
  \SetWidth{.2}
\Photon(-50,30)(-35,15){2}{4}
\Line(-35,15)(20,15)
\ArrowLine(20,15)(35,15)
\Gluon(-35,-15)(22,15){-3}{10}
\Gluon(-10,-4)(-10,15){3}{4}
\Text(0,-45)[c]{(h)}
  \SetWidth{1.5}
\ArrowLine(-35,-38)(-35,-23)
\Line(-35,-23)(-35,15)
\end{picture}}
}\]
\vspace{2.2cm}
\[\vcenter{
\hbox{
  \begin{picture}(0,0)(0,0)
\SetScale{1.}
  \SetWidth{.2}
\Photon(-50,30)(-35,15){2}{4}
\Line(-35,15)(20,15)
\ArrowLine(20,15)(35,15)
\Gluon(-35,-15)(22,15){-3}{10}
\Gluon(-10,-4)(-35,5){3}{4}
\Text(0,-45)[c]{(i)}
  \SetWidth{1.5}
\ArrowLine(-35,-38)(-35,-23)
\Line(-35,-23)(-35,15)
\end{picture}}
}
\hspace{4.cm}
\vcenter{
\hbox{
  \begin{picture}(0,0)(0,0)
\SetScale{1.}
  \SetWidth{.2}
\Photon(-50,30)(-35,15){2}{4}
\Line(-35,15)(20,15)
\ArrowLine(20,15)(35,15)
\Gluon(-35,-15)(-15,-15){-3}{2}
\Gluon(10,-15)(22,15){-3}{4}
\ArrowArc(-2,-15)(13,0,180)
\ArrowArc(-2,-15)(13,180,360)
\Text(0,-45)[c]{(j)}
  \SetWidth{1.5}
\ArrowLine(-35,-38)(-35,-23)
\Line(-35,-23)(-35,15)
\end{picture}}
}
\hspace{4.cm}
\vcenter{
\hbox{
  \begin{picture}(0,0)(0,0)
\SetScale{1.}
  \SetWidth{.2}
\Photon(-50,30)(-35,15){2}{4}
\Line(-35,15)(20,15)
\ArrowLine(20,15)(35,15)
\Gluon(-35,-15)(-15,-15){-3}{2}
\Gluon(10,-15)(22,15){-3}{4}
\Text(0,-45)[c]{(k)}
  \SetWidth{1.5}
\ArrowLine(-35,-38)(-35,-23)
\Line(-35,-23)(-35,15)
\ArrowArc(-2,-15)(13,0,180)
\ArrowArc(-2,-15)(13,180,360)
\end{picture}}
}
\hspace{4.cm}
\vcenter{
\hbox{
  \begin{picture}(0,0)(0,0)
\SetScale{1.}
  \SetWidth{.2}
\Photon(-50,30)(-35,15){2}{4}
\Line(-35,15)(20,15)
\ArrowLine(20,15)(35,15)
\Gluon(-35,-15)(-15,-15){-3}{2}
\Gluon(10,-15)(22,15){-3}{4}
\DashArrowArc(-2,-15)(13,0,180){4}
\DashArrowArc(-2,-15)(13,180,360){4}
\Text(0,-45)[c]{(l)}
  \SetWidth{1.5}
\ArrowLine(-35,-38)(-35,-23)
\Line(-35,-23)(-35,15)
\end{picture}}
}\]
\vspace*{14mm}
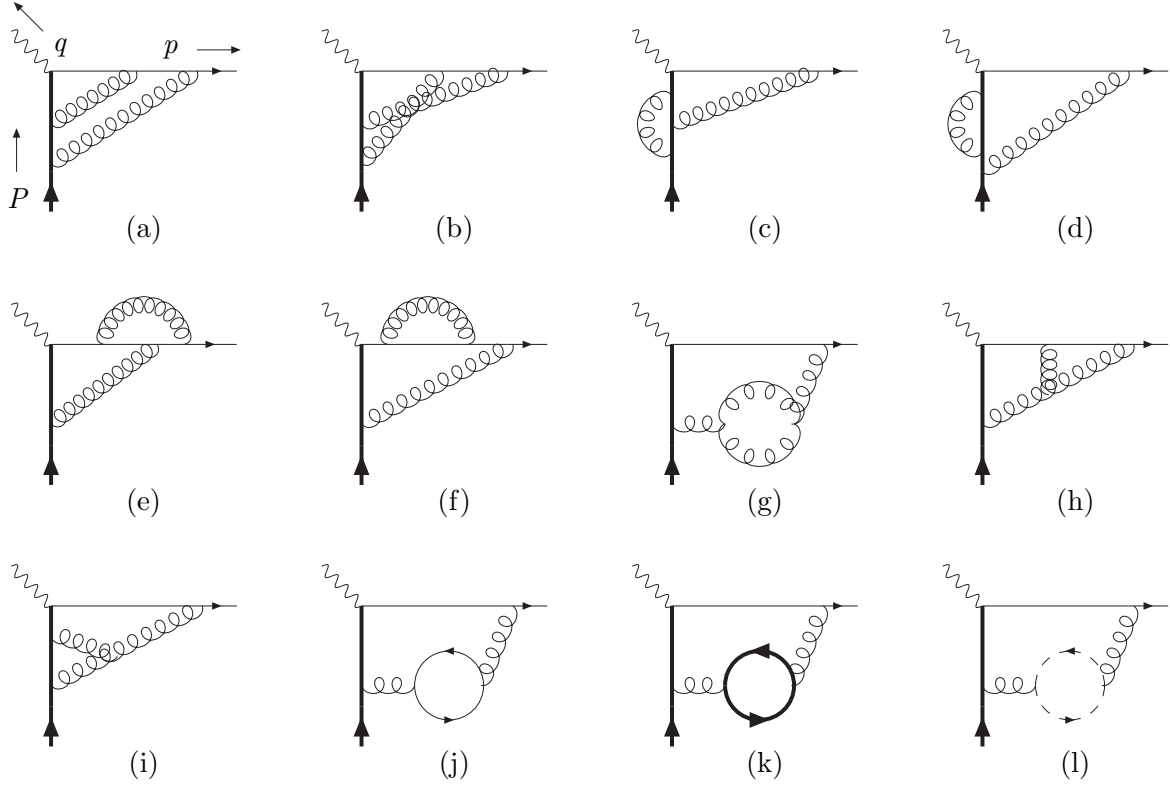
\caption{\label{fig1} Feynman Diagrams for the Two-loop QCD corrections
to the $b \to  u \, W^*$ decay process.}
\ec
\efig
%
The most general vertex correction in the Standard Model can be
described in terms of six form factors $F_i$ and $G_i$
($i=1,2,3$):
\bea 
V^{\mu}(P,p) &=&
F_1(q^2) \gamma^{\mu} + \frac{1}{2 m_b} F_2(q^2) \sigma^{\mu \nu}
q_{\nu} + \frac{i}{2 m_b} F_3(q^2) q^{\mu} + G_1(q^2) \gamma^{\mu}
\gamma_{5}  + \frac{i}{2 m_b} G_2(q^2) \gamma_{5} q^{\mu}
\nn\\
& & + \frac{i}{2 m_b} G_3(q^2) \gamma_{5} \tilde{q}^{\mu} \, ,
\label{Vmu}
\eea
where $\tilde{q}_{\mu} = P_{\mu} + p_{\mu}$. The spinors $\overline{u}(p)$
and $u(P)$, multiplying Eq.~(\ref{Vmu}) from left and right, respectively,
are not written down explicitly. We define
$\sigma^{\mu \nu} = -i/2[\gamma^\mu,\gamma^\nu]$.
Since the $u$-quark is taken as massless, only three of the above form
factors are independent. By replacing
\be
\overline{u}(p) \left[\gamma^\mu,\gamma^\nu \right]u(P) q_\nu =
2 i m_b \overline{u}(p) \gamma^\mu u(P) - 2 \tilde{q}^{\mu}\overline{u}(p) u(P)
\, ,
\ee
in Eq.~(\ref{Vmu}), we find the following relations among the form 
factors in Eq.~(\ref{Vmu}):
\be
F_2 =-G_3 \, , \qquad F_3 = -G_2 \, , \qquad F_1 + \frac{1}{2}
F_2 = G_1 \, .
\ee
Consequently, using the definitions\footnote{We employ the notation and conventions of \cite{Diagrammatica}. In particular,
in our notation $\gamma_5 = - \gamma_5^{\mbox{{\tiny bd}}}$, where
$\gamma_5^{\mbox{{\tiny bd}}}$ is the matrix commonly employed in the Bjorken-Drell notation.} $P_L = (1+\gamma_5)/2$ and
$P_R = (1-\gamma_5)/2$, we can rewrite the vertex structure as follows:
\be 
V^{\mu}(P,p) = 2 G_1(q^2) \gamma^{\mu} P_L  - \frac{i}{m_b}
G_2(q^2) P_R q^{\mu} - \frac{i}{m_b} G_3(q^2) P_R \tilde{q}^{\mu}
\, .
\ee

The form factors are expanded in powers of $\alpha_s$:
\be
G_i =\frac{i g_w}{2 \sqrt{2}} V_{ub}\left[G_i^{(0 l)} + \left(\frac{\alpha_s}{\pi} \right)G_i^{(1 l)} +
\left(\frac{\alpha_s}{\pi} \right)^2   G_i^{(2 l)} +
{\mathcal O}\left( \frac{\alpha_s^3}{\pi^3}\right) \right]\, . \label{Gexp}
\ee
The purpose of the present work is to evaluate  $G_i^{(2 l)}$ where $i=1,2,3$.
$V_{ub}$ represents the CKM matrix element and $g_w$ the weak interaction
coupling constant. With the normalization chosen
in Eq.~(\ref{Gexp}), one finds
\be
G_1^{(0 l)} = 1 \, , \qquad \mbox{and} \qquad  G_2^{(0 l)} = G_3^{(0 l)} = 0 \, .
\ee
The contribution of the virtual two-loop corrections to the hadronic tensor of
\cite{DFN} can be obtained by means of the following translation rules:
\bea
W_1^{(2 l, \mbox{{\tiny vir.}})} & = & \frac{2}{m_b^2} \left[ 2 G_1^{(2 l)} + \left( G_1^{(1 l)}
                \right)^2 \right]  , \\
W_3^{(2 l, \mbox{{\tiny vir.}})} & = & \frac{1-x}{4 m_b} \left( G_2^{(1l)} + G_3^{(1l)} \right)^2  , \\
W_4^{(2 l, \mbox{{\tiny vir.}})} & = & \frac{1}{m_b^2} \left\{ G_3^{(2 l)} \! \! +  G_2^{(2 l)} \! \!
              + G_1^{(1l)} \! \left( G_3^{(1l)} \! \! +  G_2^{(1l)} \right) + \! \frac{1\!-\!x}{4} \left[
                \left( G_3^{(1l)} \right)^2 \! \! - \! \! \left( G_2^{(1l)} \right)^2 \right] \! \right\} , \\
W_5^{(2 l, \mbox{{\tiny vir.}})} & = & \frac{2}{m_b^3} \left[ G_3^{(2l)} -  G_2^{(2l)}
              + G_1^{(1l)} \left( G_3^{(1l)} -  G_2^{(1l)} \right) + \frac{1\!-\!x}{8} \left(
                G_3^{(1l)} - G_2^{(1l)} \right)^2 \right]  .
\label{DFNrel}
\eea
Consequently, the {\it hard functions} $H_{ij}$, as defined in
\cite{Bosch:2004th}, can be extracted using the relation between Eqs.~(16) and
(17) of the same article. Note that the hadronic form factor $W_2$ does not
receive contributions from two-loop virtual corrections.

We write our analytic results in terms of the dimensionless variable
\bea
x = - \frac{q^2}{m_b^2} = \frac{M_l^2}{m_b^2} \, , \quad 0 \leq x \leq 1  \, .
\eea
In writing or results, we employ the Harmonic Polylogarithms as defined in
\cite{HPLs}; on top of the canonical weights, we introduce the weights
$-2$ and $2$, arising from the integrating factors $1/(x+2)$ and $1/(2-x)$, respectively.
Actually, only two HPLs containing the weight $2$ appear in the final result; they are
\bea
H(2;x) &=& - \ln\left( 1 - \frac{x}{2}\right) = - H(-1,1-x) + \ln(2) \, , \nn \\
H(2,1,1;x)&=& \int_0^x dt \frac{1}{2(2-t)}  \ln^2\left(1-t \right) = \nn \\
& = & - \frac{1}{2}\ln(1 - x)^2 \ln(2 - x) - \ln(1 - x) \mbox{Li}_2(-1 + x) + \mbox{Li}_3( -1 + x)
      + \frac{3}{4} \zeta\left(3\right) \, , \nn \\
& = & - H(-1,0,0,1-x) + \frac{3}{4} \zeta(3) \, .
\eea

For convenience, all the results of the paper, including the renormalized and bare form
factors, as well as the contributions of individuals diagrams, are collected in the file
{\tt SemilepFF.txt} \cite{submission} included in the arXiv submission of the present work.

\section{UV Renormalization \label{ren}}

The UV renormalization is performed by subtracting the one-loop
sub-divergencies and the two-loop over-all divergencies. We
renormalize the heavy- and light-quark wave functions and
heavy-quark mass in the {\it on-shell} ({\it OS\,}) scheme, while
the coupling constant $\alpha_S$  is renormalized in the
$\overline{\mbox{MS}}$ scheme.

Neglecting for the time being mass renormalization, the bare and renormalized form factors satisfy the relation
\be
G = Z_{2,u}^{\frac{1}{2}}
Z_{2,b}^{\frac{1}{2}} G_{\mbox{{\small bare}}} (\alpha^{\mbox{\tiny bare}}_s) \, ,
\ee
where in the functions $G$ we dropped the subscript $i=1,2,3$.

The perturbative expansion of the various quantities in the equation above is
\bea
G  &=& \frac{i g_w}{2 \sqrt{2}} V_{ub} \left(G^{(0 l )} +
a  G^{(1 l )}  + a^2G^{(2 l )} +
{\mathcal O}\left( a_0^3\right) \right) \, , \nn \\
G_{\mbox{{\small bare}}}  &=&\frac{i g_w}{2 \sqrt{2}} V_{ub}
\left(G^{(0 l )}+ a_0  G^{(1l)}_{\mbox{{\small bare}}} + a_0^2  G^{(2l)}_{\mbox{{\small bare}}}
+ {\mathcal O}\left( a_0^3\right)
\right)\, , \nn \\
Z_{2,u} &=& 1  + a_0 \delta Z^{(1l)}_{2,u} + a_0^2 \delta Z^{(2l)}_{2,u} +
{\mathcal O}\left( a_0^3\right) \, , \nn \\
Z_{2,b} &=& 1  + a_0 \delta Z^{(1l)}_{2,b} + a_0^2 \delta Z^{(2l)}_{2,b} +
{\mathcal O}\left( a_0^3\right) \, , \nn \\
a_0 &=& a\left( 1 + a \delta Z_\alpha^{(1 l)} + a^2 \delta Z_\alpha^{(2 l)} +
{\mathcal O}\left( a^3\right)  \right)   \, ,
\eea
where we defined
\be
a_0 \equiv \frac{\alpha^{\mbox{\tiny bare}}_s}{ \pi} \, , \qquad
a \equiv \frac{\alpha_s}{\pi} \, .
\ee
Therefore, the one-loop renormalized amplitude is given by
\be
G^{(1 l )} =
G^{(1 l )} _{\mbox{{\small bare}}} + \frac{1}{2} \delta Z^{(1l)}_{2,b}G^{(0 l )} \, ,
\ee
where we already took into account the fact that $\delta Z^{(1l)}_{2,u} = 0$
in the on-shell scheme.
The two-loop renormalized amplitude reads instead
\bea
G^{(2 l )}  &=&
G^{(2 l )} _{\mbox{{\small bare}}} +\left[
 \frac{1}{2}   \delta Z^{(2l)}_{2,b}  + \frac{1}{2}\delta Z^{(2l)}_{2,u}
 + \frac{1}{2} \delta Z_\alpha^{(1 l)} \delta Z^{(1l)}_{2,b}- \frac{1}{8} \left(\delta Z^{(1l)}_{2,b}\right)^2\right]G^{(0 l )}\nn  \\ &&
+\left[\frac{1}{2}  \delta Z^{(1l)}_{2,b}
+ \delta Z_\alpha^{(1 l)} \right]G^{(1 l )} _{\mbox{{\small bare}}}  \, .
\eea

To account for mass renormalization, it is sufficient to add the contribution 
of the counter term diagram in Fig.~\ref{fig2} to the r.~h.~s. of the equation
above.

The renormalization constants are the following:
\bea
\delta Z_{\alpha,\overline{\mathrm{MS}}}^{(1l)}(d) & = &
   -C(d) \,
  \frac{1}{d-4} \left( -\frac{11}{6} C_{A}
+ \frac{1}{3} T_{R} (N_l +N_h) \right)
\label{c001}
\ , \\
\delta m_{{\mathrm OS}}^{(1l)} \Bigl( d,m,\frac{\mu^2}{m^2} \Bigr)
& = &  \, m   \, C(d) \,
\left( \frac{\mu^{2}}{m^2} \right)^{(4-d)/2} \
\frac{C_{F}}{2} \frac{(d-1)}{(d-4) \, (d-3)} \ ,
\label{c003} \\
\delta Z_{2,b}^{(1l)} \Bigl(d,\frac{\mu^2}{m^2} \Bigr)
& = &  \, C(d) \,
\left( \frac{\mu^{2}}{m^2} \right)^{(4-d)/2} \
\frac{C_{F}}{2} \frac{(d-1)}{(d-4) \, (d-3)} \ ,
\label{c004} \\
\delta Z_{2,u}^{(2l)} \Bigl(d,\frac{\mu^2}{m^2} \Bigr)
& = &   C^2(d) \,
\left( \frac{\mu^{2}}{m^2} \right)^{4-d} \frac{C_F}{8} N_h
\left( -\frac{1}{2 (d-4)} - \frac{5}{24} \right)\, , \label{c005}\\
\delta Z_{2,b}^{(2l)} \Bigl(d,\frac{\mu^2}{m^2} \Bigr)
& = &  \, C^2(d) \,
\left( \frac{\mu^{2}}{m^2} \right)^{4-d} \
\frac{C_{F}}{2} \left[ C_F f_1 + C_A f_2  + \frac{1}{2} N_l f_3+ \frac{1}{2} N_h f_4 \right]  ,
\label{c006}
\eea
where $\mu$ is the renormalization scale and the constants $f_1,
\cdots ,f_4$ are \cite{vanMe}
\bea
f_1 &=& \frac{9}{8 (d-4)^2}  - \frac{51}{32 (d-4)} + \frac{433}{128}
-\frac{3}{2} \zeta_3 - \pi^2 \ln(2) -\frac{13}{16} \pi^2 + {\mathcal O}(d-4)\, , \\
f_2 &=& -\frac{11}{8 (d-4)^2} + \frac{101}{32 (d-4)} -\frac{803}{128} +
 \frac{3}{4} \zeta_3 -\frac{\pi^2}{2} \ln(2) + \frac{5}{16} \pi^2+ {\mathcal O}(d-4)
  \, , \\
f_3 &=& \frac{1}{2 (d-4)^2} - \frac{9}{8 (d-4)} +\frac{59}{32} + \frac{\pi^2}{12}+ {\mathcal O}(d-4)\, , \\
f_4 &=& \frac{1}{(d-4)^2} -\frac{19}{24 (d-4)} +\frac{1139}{288} -
\frac{\pi^2}{3} + {\mathcal O}(d-4)\, . \\
\eea
The  factor $C(d)$ is
\be
C(d) = \left(4 \pi\right)^{(4-d)/2} \Gamma\left(3- \frac{d}{2} \right) \, .
\label{Cd}
\ee

After UV renormalization, the vertex form factors still contain poles in
$1/(d-4)$, which are associated to soft and collinear singularities.

\begin{figure}
\vspace*{10mm}
\bc
\[ \vcenter{
\hbox{
  \begin{picture}(0,0)(0,0)
\SetScale{1.}
  \SetWidth{.2}
\Photon(-50,30)(-35,15){2}{4}
\Line(-35,15)(20,15)
\ArrowLine(20,15)(35,15)
\Gluon(-35,-20)(20,15){-3}{10}
\CCirc(-35,-4){4}{Black}{Black}
\Text(10,-20)[c]{$- \frac{\delta m_{{\mathrm OS}}^{(1l)}}{m}$}
  \SetWidth{1.5}
\ArrowLine(-35,-38)(-35,-23)
\Line(-35,-23)(-35,15)
\end{picture}}
}\]
\vspace*{10mm}
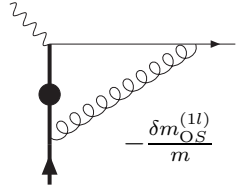
\caption{\label{fig2} Mass-renormalization counter-term.}
\ec
\efig
%

\section{One-Loop Form Factors \label{1lFF}}

In this section we collect the analytic expression of the one-loop renormalized
form factors defined in Eq.~(\ref{Gexp}). In the formulas below,
$C_F=(N_c^2-1)/2N_c$ is the Casimir operator of the fundamental representation
of $SU(N_c)$, where $N_c$ is the number of colors (in the SM $N_c=3$).

The form factor $G_1^{(1 l)}$ is given by
\bea
G_1^{(1 l)} &=& C(d) \left( \frac{\mu^2}{m^2}\right)^{\frac{4-d}{2}}
C_F \sum_{i = -2}^{1} G_1^{(1 l,i)} (d-4)^i + {\mathcal O}\left((d-4)^2 \right)
\, ,
\eea
where  the first four coefficients in the expansion in $(d-4)$ are
\bea
 G_1^{(1 l,-2)} &=& -1 \, , \nn \\
 G_1^{(1 l,-1)} &=& \frac{5}{4} + H(1;x) \, , \nn \\
 G_1^{(1 l,0)} &=& - \frac{3}{2}
                   + \frac{1 - 3 x}{4 x} H(1;x)
           - \frac{1}{2} H(0, 1;x)
           - H(1, 1;x) \, , \nn \\
G_1^{(1 l, 1)} &=& \frac{3}{2}
                  - \frac{1 - 2 x}{2 x} H(1;x)
                  - \frac{1 - 3 x}{8 x} H(0, 1;x)
          - \frac{1 - 3 x}{4 x} H(1, 1;x)
          + \frac{1}{4} H(0, 0, 1;x)  \nn \\
& &
          + \frac{1}{2} H(0, 1, 1;x)
          + \frac{1}{2} H(1, 0, 1;x)
          + H(1, 1, 1;x)
\, .
\eea

The form factor $G_2^{(1 l)}$ is
\bea
G_2^{(1 l)} & = & C(d) \left( \frac{\mu^2}{m^2}\right)^{\frac{4-d}{2}}
C_F \sum_{i = 0}^{1} G_2^{(1 l,i)} (d-4)^i+ {\mathcal O}\left((d-4)^2 \right) \, ,
\eea
with
\bea
G_2^{(1 l,0)} & = & \frac{1}{x}
                   - \frac{2 - 3 x}{2 x^2} H(1;x)
\, , \nn \\
G_2^{(1 l,1)} & = & - \frac{1}{x}
                    + \frac{1 - 3 x}{2 x^2} H(1;x)
            + \frac{2 - 3 x}{4 x^2} H(0, 1;x)
            + \frac{2 - 3 x}{2 x^2} H(1, 1;x)
\, .
\eea

Finally, the form factor $G_3^{(1 l)}$ is
\bea
G_3^{(1 l)} & = & C(d) \left( \frac{\mu^2}{m^2}\right)^{\frac{4-d}{2}}
C_F \sum_{i = 0}^{1} G_2^{(1 l,i)} (d-4)^i + {\mathcal O}\left((d-4)^2 \right) \, ,
\eea
where
\bea
G_3^{(1 l,0)} & = & - \frac{1}{2 x} H(1;x) \, , \nn \\
G_3^{(1 l,1)} & = &   \frac{1}{x} H(1;x)
                    + \frac{1}{2 x} H(1,1;x)
                + \frac{1}{4 x} H(0,1;x) \, .
\eea

Note that the IR poles of the form factor $G_1^{(1 l)}$ exponentiate. This means that
from the ${\mathcal O}(\alpha_S^2)$ expansion of the form factor
\be
{\mathcal F} = \exp{\{G_1^{(1 l)}\}} \, ,
\ee
we can predict exactly the $1/(d-4)^4$ and $1/(d-4)^3$ poles of the $C_F^2$ part
of the two-loop form factor $G_1^{(2l)}$ (Eqs.~(\ref{G1p4},\ref{G1p3}) below).
Moreover, exponentiating also the finite part of $G_1^{(1 l)}$, the double pole
of Eq.~(\ref{G1p2}) is exactly recovered.

\section{Two-Loop Form Factors \label{2lFF}}

In this section we collect the analytic expression of the two-loop renormalized
form factors defined in Eq.~(\ref{Gexp}). In the expressions below, $C_A$ is the
Casimir operator of the adjoint representation of $SU(N_c)$, $C_A=N_c$, $T_R$ is
the normalization factor of the color matrices, $T_R=1/2$, $N_l$ is the number
of massless quarks in the theory, and $N_h$ is the number of quarks of mass
$m_b$. Therefore, for the decay $b \to u \, W^*$ in the SM, $N_l=4$, and 
$N_h =1$.
In the finite part of the form factors given below, the constant ${\mathcal K}$
is a rational number.  Its numerical value is ${\mathcal K}=3.32812 \pm
0.00002$,  and its analytical value is likely to be ${\mathcal K}=213/64$.
We observe that the following formulas involve HPLs of argument $x$ and of
maximal weight 4. If desired, the  HPLs appearing in the equations below can all
be rewritten in terms of product of Nielsen Polylogarithms of more complicated
argument. Because of the chosen renormalization scheme, our results depend on
the renormalization scale $\mu$. In the formulas below, we employ the following
notation:
\be
\ln\left(\frac{\mu^2}{m_b^2} \right) \equiv L_\mu \, .
\ee

 The form factor $G_1^{(2 l)}$ can be written as
\bea
G_1^{(2 l)} &=& C^2(d) \left( \frac{\mu^2}{m^2}\right)^{4-d}
C_F \sum_{i = -4}^{0} G_1^{(2 l,i)} (d-4)^i + {\mathcal O}\left(d-4 \right)\, ,
\eea
where the coefficient of the expansion in $(d-4)$ (up to the finite term) are
\bea
G_1^{(2 l,-4)} & = &  C_F  \frac{1}{2}
\, ,
\label{G1p4} \\
G_1^{(2 l,-3)} & = & C_F \biggl[ - \frac{5}{4} - H(1;x)
                         \biggr]
           - C_A \frac{11}{8}
           + T_R N_l  \frac{1}{2}
\, ,
\label{G1p3} \\
G_1^{(2 l,-2)} & = &  C_F \biggl[  \frac{73}{32}
                                 - \frac{1 \! - \! 8 x}{4 x} H(1;x)
                 + \frac{1}{2} H(0, 1;x)
                 + 2 H(1, 1;x)
                         \biggr]
           + C_A \biggl[   \frac{49 \! - \! 66 L_{\mu} \! + \! 9 \zeta(2)}{72}  \nn\\
& &
                         + \frac{11}{12} H(1;x)
                         \biggr]
           + T_R N_l \biggl[ \frac{-5 + 6 L_{\mu}}{18}
                         - \frac{1}{3} H(1;x)
                         \biggr]
           + T_R N_h \biggl[  \frac{1}{3} L_{\mu}
                         \biggr]
\, ,
\label{G1p2} \\
G_1^{(2 l,-1)} & = & C_F \biggl[ - \frac{3(71 + 8 \zeta(2) - 16 \zeta(3))}{64}
                                 + \frac{13 - 55 x}{16 x} H(1;x)
                 + \frac{1 - 8 x}{8 x} H(0, 1;x)  \nn\\
& &
                 + \frac{3 \! - \! 14 x}{4 x} H(1, 1;x)\!
                 - \! \frac{1}{4} H(0, 0, 1;x) \!
                 - \! \frac{3}{2} H(0, 1, 1;x) \!
                 - \! H(1, 0, 1;x) \!
                 - \! 4 H(1, 1, 1;x)
                         \biggr]  \nn\\
& &
           + C_A \biggl[   \frac{1549 \! + \! 1980 L_{\mu} \! - \! 396 L_{\mu}^2 \! + \! 972 \zeta(2) \!
                         - \! 1188 \zeta(3)}{1728}
                         + \frac{67 \! + \! 66 L_{\mu} \! - \! 18 \zeta(2)}{72} H(1;x)
                         \biggr]  \nn\\
& &
           + T_R N_l \biggl[  \frac{-125 - 180 L_{\mu} + 36 L_{\mu}^2 - 108 \zeta(2)}{432}
                             - \frac{5 + 6 L_{\mu}}{18}  H(1;x)
                         \biggr]  \nn\\
& &
           + T_R N_h \biggl[  \frac{-5 L_{\mu} + L_{\mu}^2 - \zeta(2)}{12}
                             - \frac{L_{\mu}}{3} H(1;x)
                         \biggr]
\, , \nn \\
G_1^{(2 l,0)}  & = &  C_F \biggl[ \frac{1}{1280 (x-1)^3} (
                        6635 (x-1)^3 + (80 (x-1) (-59 - 62 x + 3 x^2) -
                        480 \ln{(2)} (10   \nn\\
& &
                                        - 22 x + 5 x^2 + 4 x^3)) \zeta(2) -
                        16 (261 - 64 {\mathcal K} + 41 x + 192 {\mathcal K} x + 405 x^2 - 192 {\mathcal K} x^2  \nn\\
& &
                                        - 99 x^3 + 64 {\mathcal K} x^3) \zeta^2(2) + 40 (52 - 72 x + 15 x^2 + 14 x^3) \zeta(3) )
                             + \frac{1}{4 (x-1)^2 x} (\zeta(2)  \nn\\
& &
                                         - 3 x \zeta(2) + x^2 \zeta(2) + 5 x^3 \zeta(2))  H(-1;x)
                 + \frac{1}{32 (x-1)^2 x} (-49 + 251 x - 355 x^2   \nn\\
& &
                                                 + 153 x^3
                                 + 12 \zeta(2) - 160 x \zeta(2) - 148 x^2 \zeta(2) + 24 x^3 \zeta(2)
                         - 16 x \zeta(3) + 32 x^2 \zeta(3)  \nn\\
& &
                                                 - 16 x^3 \zeta(3)) H(1;x)
                 + \frac{3 (-2 \zeta(2) - 2 x \zeta(2) + x^2 \zeta(2))}{8 (x-1)^3} H(2;x) \nn\\
& &
                 + \frac{\zeta(2) - 5 x \zeta(2) + 3 x^2 \zeta(2) - x^3 \zeta(2)}{2 (x-1)^3}  H(0, -1;x)
                 + \frac{1}{32 (x-1)^3 x} (15 - 106 x  \nn\\
& &
                                              + 248 x^2 - 138 x^3
                              - 19 x^4 + 72 x \zeta(2) + 152 x^2 \zeta(2) + 48 x^3 \zeta(2)) H(0, 1;x)  \nn\\
& &
                     + \frac{25 - 134 x + 59 x^2}{16 (x-1) x}  H(1, 1;x)
                 + \frac{1 - 3 x + x^2 + 5 x^3}{2 (x-1)^2 x}  H(-1, 0, 1;x) \nn\\
& &
                 - \frac{1 \! + \! 27 x \! - \! 9 x^2 \! - \! 25 x^3 \! + \! 12 x^4}{16 (x-1)^3 x}  H(0, 0, \! 1;x)\!
                 + \! \frac{5 \! - \! 59 x\!  + \! 83 x^2\!  -\!  56 x^3 \! + \! 30 x^4}{8 (x-1)^3 x}  H(0, \! 1, \! 1;x) \nn\\
& &
                 - \frac{1 \! - \! x \! + \! 21 x^2 \! - \! 7 x^3}{4 (x-1)^2 x} H(1, 0, 1;x)
                 - \frac{7 \! - \! 26 x}{4 x}  H(1, 1, 1;x)
                 - \frac{2 \! + \! 2 x \! - \! x^2}{8 (x-1)^3}  H(2, 1, 1;x) \nn\\
& &
                 + \frac{1 - 5 x + 3 x^2 - x^3}{(x-1)^3}  H(0, -1, 0, 1;x)
                 - \frac{3 + 11 x - 5 x^2 + 3 x^3}{8 (x-1)^3} H(0, 0, 0, 1;x) \nn\\
& &
                 - \frac{9 \! + \! 9 x \! + \! 13 x^2 \! - \! 3 x^3}{4 (x-1)^3}  H(0, \! 0, \! 1,\!  1;x)\!
                 + \! \frac{18 x \! - \! 7 x^2 \! + \! 3 x^3}{4 (x-1)^3}  H(0, \! 1,\!  0, \! 1;x)
                 + \! 3 H(1, \! 0, \! 1, \! 1;x) \nn\\
& &
                 + \! \frac{7}{2} H(0, \! 1, \! 1, \! 1;x) \!
                 - \! \frac{1}{2} H(1, 0, 0, \! 1;x) \!
                 + \! 2 H(1, \! 1, 0, \! 1;x) \!
                 +\!  8 H(1,\!  1, \! 1, \! 1;x) \!
                         \biggr] \nn\\
& &
           + C_A \biggl[   \frac{ 29700 L_{\mu}^2 \! -447185 \! - \! 142560 L_{\mu} \! - \! 3960 L_{\mu}^3 }{103680}
                +  \frac{1}{103680 (x\! \! -\! \!  1)^3} ((11880 L_{\mu} (x\! \! -\! 1)^3  \nn\\
& &
                        - 180 (x-1) (517 + 982 x +
                        913 x^2) + 19440 \ln{(2)} (10 - 22 x + 5 x^2 + 4 x^3)) \zeta(2)  \nn\\
& &
                                                + 648 (71 - 64 {\mathcal K} - 631 x + 192 {\mathcal K} x + 159 x^2 - 192 {\mathcal K} x^2 -
                        125 x^3 + 64 {\mathcal K} x^3) \zeta^2(2)  \nn\\
& &
                                                + 180 (-698 + 1338 x - 825 x^2 + 104 x^3) \zeta(3) )
                             - \frac{(1 - 3 x + x^2 + 5 x^3 )\zeta(2)}{8 (x-1)^2 x}  H(-1;x) \nn\\
& &
                 + \frac{1}{864 (x-1)^2 x} (807 \! + \! 198 L_{\mu} \! - \! 4159 x \! - \! 990 L_{\mu} x \!
                            + \! 198 L_{\mu}^2 x \! + \! 5897 x^2 \! + \! 1386 L_{\mu} x^2  \nn\\
& &
                                            - 396 L_{\mu}^2 x^2 \!
                        - \! 2545 x^3 \! - \! 594 L_{\mu} x^3 \! + \! 198 L_{\mu}^2 x^3 \!
                        - \! ( 216
                        + 1368 x + 342 x^2
                                            + 1206 x^3 ) \zeta(2)  \nn\\
& &
                                            + ( 756 x - 1512 x^2 + 756 x^3) \zeta(3)  ) H(1;x)
                     + \frac{3 (2 + 2 x - x^2) \zeta(2)}{16 (x-1)^3}  H(2;x) \nn\\
& &
                 - \frac{(1 - 5 x + 3 x^2 - x^3) \zeta(2)}{4 (x-1)^3} H(0, -1;x)
                 + \frac{1}{144 (x-1)^3 x} (-33 + 260 x + 66 L_{\mu} x  \nn\\
& &
                                           - \! 384 x^2 \!
                           - \! 198 L_{\mu} x^2 \! + \! 273 x^3 \! + \! 198 L_{\mu} x^3 \! - \! 116 x^4 \!
                       - \! 66 L_{\mu} x^4 \! - \! 18 x \zeta(2) \! + \! 468 x^2 \zeta(2)  \nn\\
& &
                                           + \! 72 x^4 \zeta(2)) H(0, 1;x)
                 + \frac{1}{144 (x-1) x} (-39 \! + \! 235 x \! + \! 132 L_{\mu} x \! - \! 349 x^2 \! - \! 132 L_{\mu} x^2  \nn\\
& &
                                                    - 72 x \zeta(2)
                                    + 72 x^2 \zeta(2)) H(1, 1;x)
                 - \frac{1 - 3 x + x^2 + 5 x^3}{4 (x-1)^2 x} H(-1, 0, 1;x) \nn\\
& &
                 - \frac{8 + 36 x - 33 x^2 - 20 x^3}{48 (x-1)^3}  H(0, 0, 1;x)
                 + \frac{2 - 168 x + 219 x^2 - 62 x^3}{48 (x-1)^3} H(0, 1, 1;x) \nn\\
& &
                 - \frac{47 - 49 x + 44 x^2}{48 (x-1)^2} H(1, 0, 1;x)
                 - \frac{11}{6} H(1, 1, 1;x)
                 + \frac{2 + 2 x - x^2}{16 (x-1)^3} H(2, 1, 1;x) \nn\\
& &
                 - \frac{1 - 5 x + 3 x^2 - x^3}{2 (x-1)^3} H(0, -1, 0, 1;x)
                 - \frac{1 + 10 x + 4 x^2}{8 (x-1)^3} H(0, 0, 0, 1;x) \nn\\
& &
                 - \frac{1 + 2 x + 4 x^2}{4 (x-1)^3} H(0, 0, 1, 1;x)
                 + \frac{1 + 2 x + 4 x^2}{8 (x-1)^3} H(0, 1, 0, 1;x)
                 + \frac{1}{2} H(1, 0, 0, 1;x)
                         \biggr]  \nn\\
& &
           + T_R N_l \biggl[ \frac{1}{5184} (6629 \! + \! 2592 L_{\mu} \! - \! 540 L_{\mu}^2 \! + \! 72 L_{\mu}^3 \!
                                  + \! 3420 \zeta(2) \! - \! 216 L_{\mu} \zeta(2)
                                  + 720 \zeta(3))  \nn\\
& &
                         - \frac{57 \! + 18 L_{\mu} \! - \! 209 x \! - \! 54 L_{\mu} x \! + \! 18 L_{\mu}^2 x \!
                     - \! 90 x \zeta(2)}{216 x}  H(1;x)
                     \! - \frac{3 \! - \! 19 x \! - \! 6 L_{\mu} x}{36 x} H(0, 1;x) \nn\\
& &
                     - \frac{3 - 19 x - 6 L_{\mu} x}{18 x}  H(1, 1;x)
                     + \frac{1}{6} H(0, 0, 1;x)
                     + \frac{1}{3} H(0, 1, 1;x)
                     + \frac{1}{3} H(1, 0, 1;x)  \nn\\
& &
                     + \frac{2}{3} H(1, 1, 1;x)
                         \biggr]
           + T_R N_h \biggl[ \frac{1}{2592 (x-1)^3} (-1111 - 1296 L_{\mu} + 270 L_{\mu}^2 - 36 L_{\mu}^3  \nn\\
& &
                                           + 7869 x
                           + 3888 L_{\mu} x - 810 L_{\mu}^2 x +
                           108 L_{\mu}^3 x - 14709 x^2 - 3888 L_{\mu} x^2 + 810 L_{\mu}^2 x^2  \nn\\
& &
                                           - 108 L_{\mu}^3 x^2 + 7951 x^3 + 1296 L_{\mu} x^3 - 270 L_{\mu}^2 x^3 +
                           36 L_{\mu}^3 x^3 - 414 \zeta(2) + 108 L_{\mu} \zeta(2)  \nn\\
& &
                                           - 5670 x \zeta(2) -
                           324 L_{\mu} x \zeta(2) + 9126 x^2 \zeta(2) + 324 L_{\mu} x^2 \zeta(2) - 738 x^3 \zeta(2)  \nn\\
& &
                                           - 108 L_{\mu} x^3 \zeta(2) + 504 \zeta(3) + 1080 x \zeta(3) + 1512 x^2 \zeta(3) -
                           504 x^3 \zeta(3))  \nn\\
& &
                                 + \frac{1}{216 (x-1)^2 x} (-57 - 18 L_{\mu} - 89 x + 90 L_{\mu} x - 18 L_{\mu}^2 x
                              + 73 x^2  \nn\\
& &
                                              - 126 L_{\mu} x^2 + 36 L_{\mu}^2 x^2 + 265 x^3 + 54 L_{\mu} x^3
                          - 18 L_{\mu}^2 x^3 + 18 x \zeta(2) - 36 x^2 \zeta(2)  \nn\\
& &
                                              + \! 18 x^3 \zeta(2)) H(1;x)\!
                     + \frac{3 \! + \! 8 x \! - \! 24 x^3 \! - \! 19 x^4 \!  -\! ( 6 x \! - \! 18 x^2 \! + \! 18 x^3
                              \! - \! 6 x^4) L_{\mu}}{36 (x-1)^3 x} H(0, 1;x)  \nn\\
& &
                         + \frac{1}{3} L_{\mu} H(1, 1;x)
                     - \frac{1 + 3 x + 3 x^2 - x^3}{6 (x-1)^3} H(0, 0, 1;x)
                         \biggr]
\, .
\label{G12l}
\eea

The form factor $G_2^{(2 l)}$ is
\bea
G_2^{(2 l)} &=& C^2(d) \left( \frac{\mu^2}{m^2}\right)^{4-d}
C_F \sum_{i = -2}^{0} G_2^{(2 l,i)} (d-4)^i + {\mathcal O}\left( d-4 \right) \, ,
\eea
where
\bea
G_2^{(2 l,-2)} &=& C_F \biggl[  - \frac{1}{x}
                                + \frac{2 - 3 x}{2 x^2} H(1;x)
                         \biggr]
\, , \\
G_2^{(2 l,-1)} &=& C_F \biggl[  \frac{9}{4 x}
                               - \frac{7 (2 - 5 x)}{8 x^2}  H(1;x)
                   - \frac{(2 - 3 x)}{4 x^2} H(0, 1;x)
                   - \frac{3 (2 - 3 x)}{2 x^2} H(1, 1;x)
                         \biggr]
\, , \\
G_2^{(2 l,0)} &=& C_F \biggl[  \frac{1}{80 (x-1)^4 x} (-310 (x-1)^4 + (60 \ln{(2)} x (-38 + 58 x - 40 x^2 + 11 x^3) \nn\\
& &
                        - 20 (x-1) (10 - 120 x - 79 x^2 + 12 x^3)) \zeta(2) +
                        16 x (125 + 103 x) \zeta^2(2) \nn\\
& &
                                        - 5 x (-30 + 110 x - 80 x^2 + 27 x^3) \zeta(3) )
                          + \frac{(2 \! - \! 9 x \! - \! 5 x^2 \! + \! 3 x^3\! - \! 3 x^4) \zeta(2)}{2 (x-1)^3 x^2}  H(-1;x) \nn\\
& &
                  + \frac{1}{16 (x-1)^3 x^2} (-32 + 195 x - 397 x^2 + 337 x^3 - 103 x^4 + 24 \zeta(2) -
                                              12 x \zeta(2)  \nn\\
& &
                                            + \! 844 x^2 \zeta(2) \! - \! 76 x^3 \zeta(2) \! + \! 36 x^4 \zeta(2)) H(1;x)
                  + \frac{3 (30 \! - \! 34 x \! + \! 16 x^2 \! - \! 3 x^3) \zeta(2)}{4 (x-1)^4} H(2;x)  \nn\\
& &
                  + \frac{2 (2 + x) \zeta(2)}{(x-1)^4}  H(0, -1;x)
                  + \frac{1}{16 (x-1)^4 x^2}(26 - 69 x - 68 x^2 - 58 x^3 + 166 x^4  \nn\\
& &
                                         + \! 3 x^5 - \! (  448 x^2 \!
                             + \! 368 x^3 ) \zeta(2)) H(0, 1;x)
                  + \frac{8 \! - \! 18 x \! + \! 37 x^2 \! + \! 172 x^3 \! - \! 49 x^4}{8 (x-1)^2 x^3} H(1, 1;x) \nn\\
& &
                  + \frac{2 - 9 x - 5 x^2 + 3 x^3 - 3 x^4}{(x-1)^3 x^2} H(-1, 0, 1;x)
                  + \frac{30 - 34 x + 16 x^2 - 3 x^3}{4 (x-1)^4} H(2, 1, 1;x) \nn\\
& &
                  - \frac{2 - 27 x + 4 x^2 - 48 x^3 + 66 x^4 - 15 x^5}{8 (x-1)^4 x^2} H(0, 0, 1;x)
                  + \frac{7 (2 - 3 x)}{2 x^2} H(1, 1, 1;x)  \nn\\
& &
                  + \frac{10 \! - \! 39 x \! + \! 234 x^2 \! - \! 276 x^3 \! + \! 86 x^4 \!
                          - \! 24 x^5}{4 (x-1)^4 x^2} H(0, 1, 1;x)
                  + \frac{4 (2 + x)}{(x-1)^4} H(0, -1, 0, 1;x)  \nn\\
& &
                  - \frac{2 - 13 x - 22 x^2 - 12 x^3 + 3 x^4}{2 (x-1)^3 x^2} H(1, 0, 1;x)
                  + \frac{(4 + 5 x)}{(x-1)^4} H(0, 0, 0, 1;x)  \nn\\
& &
                  + \frac{6 (4 + 3 x)}{(x-1)^4}  H(0, 0, 1, 1;x)
                  - \frac{3 (4 + 3 x)}{(x-1)^4} H(0, 1, 0, 1;x)
                         \biggr]  \nn\\
& &
           + C_A \biggl[ \frac{1}{1440 (x-1)^4 x} (1320 L_{\mu} (x-1)^4 + 20 (x-1)^3 (-269 + 242 x)  \nn\\
& &
                                           + (540 (x-1) (4 + 50 x + 5 x^2 + 8 x^3) +
                            540 \ln{(2)} x (38 - 58 x + 40 x^2 - 11 x^3)) \zeta(2)  \nn\\
& &
                                           + 36 x (364 + 317 x + 108 x^2) \zeta^2(2) +
                       45 x (-30 + 110 x - 80 x^2 + 27 x^3) \zeta(3)) \nn\\
& &
                             - \frac{(2 \! - \! 9 x \! - \! 5 x^2 \! + \! 3 x^3 \! - \! 3 x^4) \zeta(2)}{4 (x-1)^3 x^2}  H(-1;x)
                 + \frac{1}{144 (x-1)^3 x^2} (406 \! + \! 132 L_{\mu} \! - \! 2067 x  \nn\\
& &
                                           - 594 L_{\mu} x \! + \! 3603 x^2 \!\! + \!
                           990 L_{\mu} x^2\! \! -\!  2629 x^3 \!\! - \! 726 L_{\mu} x^3 \!\! + \! 687 x^4 \! + \!
                           198 L_{\mu} x^4\!  - \! ( 144 \! - \! 864 x \nn\\
& &
                                            - 1224 x^2 \! - \! 1242 x^3
                       + 54 x^4 ) \zeta(2)) H(1;x)
                             - \frac{3 (30 - 34 x + 16 x^2 - 3 x^3) \zeta(2)}{8 (x-1)^4}  H(2;x) \nn\\
& &
                 - \frac{(2 + x) \zeta(2)}{(x-1)^4}  H(0, -1;x)
                 + \frac{1}{24 (x-1)^4 x^2} (-22 + 145 x - 300 x^2 + 109 x^3
                          + 59 x^4  \nn\\
& &
                                 + 9 x^5 \! - \! ( 240 x^2 \! + \! 210 x^3 \! + \! 72 x^4 ) \zeta(2)) H(0, 1;x)
                             - \frac{26 \! - \! 151 x \! + \! 14 x^2 \! - \! 42 x^3}{24 (x-1)^2 x^2} H(1, 1;x)  \nn\\
& &
                 - \frac{2 \! - \! 9 x \! - \! 5 x^2 \! + \! 3 x^3 \! - \! 3 x^4}{2 (x-1)^3 x^2}  H(-1, 0, 1;x)
                 + \frac{4 \! + \! 76 x \! - \! 96 x^2 \! + \! 16 x^3 \! - \! 9 x^4}{8 (x-1)^4 x} H(0, 0, 1;x)  \nn\\
& &
                 + \frac{8 + 58 x - 24 x^2 - 36 x^3 + 3 x^4}{8 (x-1)^4 x} H(0, 1, 1;x)
                 + \frac{4 + 24 x + 11 x^2 + 3 x^3}{8 (x-1)^3 x} H(1, 0, 1;x)  \nn\\
& &
                 - \frac{30 - 34 x + 16 x^2 - 3 x^3}{8 (x-1)^4} H(2, 1, 1;x)
                 + \frac{24 + 17 x + 4 x^2}{4 (x-1)^4} H(0, 0, 0, 1;x)  \nn\\
& &
                 - \frac{2 (2 + x)}{(x-1)^4} H(0, -1, 0, 1;x)
                 + \frac{8 + 9 x + 4 x^2}{4 (x-1)^4} ( 2 H(0, 0, 1, 1;x) - H(0, 1, 0, 1;x) )\!
                         \biggr]  \nn\\
& &
           + T_R N_l \biggl[ \! -  \frac{19 \! + \! 6 L_{\mu}}{18 x} \!
                             + \! \frac{26 \! - \! 51 x \! + \! 6(\! 2 \! - \! 3 x) L_{\mu}}{36 x^2}  H(\! 1;x)\!\!
                     + \!\! \frac{2\!  - \! 3 x}{6 x^2} (\! H(\! 0, \! 1;x) \! +\!  2 H(\! 1, \! 1;x) )
                         \! \biggr]  \nn\\
& &
           + T_R N_h \biggl[  \frac{1}{18 (x-1)^4 x} (-19 - 6 L_{\mu} - 164 x + 24 L_{\mu} x +
                            393 x^2 - 36 L_{\mu} x^2 - 218 x^3  \nn\\
& &
                                                + \! 24 L_{\mu} x^3 \! +\!  8 x^4 \! - \!
                            6 L_{\mu} x^4 \! + \! (252 x - 300 x^2 \! + \! 84 x^3 \!
                            - \! 36 x^4 ) \zeta(2) - ( 72 x + 36 x^2 ) \zeta(3))  \nn\\
& &
                                 + \frac{26 - 223 x - 124 x^2 - 51 x^3 + ( 12 - 42 x
                         + 48 x^2 - 18  x^3 )L_{\mu} }{36 (x-1)^2 x^2} \, H(1;x) \nn\\
& &
                         - \frac{2 - 9 x - 21 x^2 - 13 x^3 - 3 x^4}{6 (x-1)^3 x^2}  H(0, 1;x)
                     + \frac{2 (2 + x)}{(x-1)^4} H(0, 0, 1;x)
                         \biggr]
\, .
\label{G22l}
\eea

The form factor $G_3^{(2 l)}$ can be written as
\bea
G_3^{(2 l)} &=& C^2(d) \left( \frac{\mu^2}{m^2}\right)^{4-d}
C_F \sum_{i = -2}^{0} G_3^{(2 l,i)} (d-4)^i  + {\mathcal O}\left(d-4
\right)\, ,
\eea
with
\bea
G_3^{(2 l,-2)} &=&  C_F \biggl[  \frac{1}{2x} H(1;x)
                         \biggr]
\, , \\
G_3^{(2 l,-1)} &=&  C_F \biggl[  - \frac{13}{8 x} H(1;x)
                                 - \frac{1}{4 x} ( H(0, 1;x) + 6 H(1, 1;x) )
                         \biggr]
\, , \\
G_3^{(2 l,0)} &=& C_F \biggl[  \frac{1}{80 (x-1)^4} ((-60 (x-1) x (57 \! +\!  2 x) - 60 \ln{(2)} (-22 \! + \! 30 x \! - \! 26 x^2 \! +
                            \! 9 x^3)) \zeta(2)  \nn\\
& &
                                                        -\!  48 (9 \! + \! 49 x \! + \! 18 x^2) \zeta^2(2) \! - \!
                            5 (22 \! - \! 90 x \! + \! 82 x^2 \! - \! 41 x^3) \zeta(3)) \!
                 - \! \frac{1}{16 (x\! -\! 1)^3 x}(51 \!\!  - \! \! 157 x  \nn\\
& &
                                                       + 161 x^2 - 55 x^3 - ( 12 - 364 x - 436 x^2 - 28 x^3) \zeta(2)) H(1;x)  \nn\\
& &
                         + \frac{(1 + 5 x + 5 x^2 + x^3) \zeta(2)}{2 (x-1)^3 x} H(-1;x)
                 + \frac{3 (-14 + 6 x - 2 x^2 + x^3) \zeta(2)}{4 (x-1)^4}  H(2;x) \nn\\
& &
                 - \frac{6 x \zeta(2)}{(x-1)^4} H(0, -1;x)
                 + \frac{1}{16 (x\! -\! 1)^4 x} (11 \! - \! 72 x \! + \! 366 x^2 \! -\!  260 x^3 \! - \! 45 x^4 \!
                             + \! 96 x \zeta(2)  \nn\\
& &
                                         + 528 x^2 \zeta(2) + 192 x^3 \zeta(2)) H(0, 1;x)
                 - \frac{1 + 36 x + 8 x^2 - 24 x^3 - 3 x^4}{8 (x-1)^4 x} H(0, 0, 1;x)  \nn\\
& &
                 + \frac{25 \! - \! 174 x \! - \! x^2}{8 (x-1)^2 x} H(1, 1;x)
                 + \frac{1 \! + \! 5 x \! + \! 5 x^2 \! + \! x^3}{(x-1)^3 x} H(-1, 0, 1;x)
                 + \frac{7}{(2 x)} H(1, 1, 1;x)  \nn\\
& &
                 + \frac{5 - 78 x + 4 x^2 + 70 x^3 + 8 x^4}{4 (x-1)^4 x} H(0, 1, 1;x)
                 - \frac{1 + 12 x + 32 x^2 - 3 x^3}{2 (x-1)^3 x} H(1, 0, 1;x) \nn\\
& &
                 - \frac{14 - 6 x + 2 x^2 - x^3}{4 (x-1)^4} H(2, 1, 1;x)
                 - \frac{2 (2 + 15 x + 4 x^2)}{(x-1)^4} H(0, 0, 1, 1;x)  \nn\\
& &
                 - \! \frac{2 \! + \! 3 x \! + \! 4 x^2}{(x\! -\! 1)^4} H(0, \! 0, \! 0, \! 1;x) \!
                 - \! \frac{12 x}{(x\! -\! \! 1)^4} H(0, \! -1, \! 0, \! 1;x) \!
                 + \! \frac{2 \! + \! 15 x \! + \! 4 x^2}{(x\! -\! 1)^4} H(0, \! 1, \! 0, \! 1;x)\!
                         \biggr]   \nn\\
& &
           + C_A \biggl[  \frac{1}{160 (x\! -\! 1)^4} (60 (x\! -\! 1)^3 + (-20 (x\! -\! 1) (56 + 105 x + 40 x^2) +
                            60 \ln{(2)} (-22  \nn\\
& &
                                                        + 30 x - 26 x^2 + 9 x^3)) \zeta(2) -
                            12 (18 + 137 x + 108 x^2) \zeta^2(2) -
                            5 (-22 + 90 x - 82 x^2  \nn\\
& &
                                                        + 41 x^3) \zeta(3))
                     - \frac{(1 + 5 x + 5 x^2 + x^3) \zeta(2)}{4 (x-1)^3 x} H(-1;x)
                 + \frac{1}{144 (x-1)^3 x} (335 + 66 L_{\mu}  \nn\\
& &
                                           - 843 x - 198 L_{\mu} x + 681 x^2 + 198 L_{\mu} x^2 -
                           173 x^3 - 66 L_{\mu} x^3 - (72 + 468 x + 2466 x^2  \nn\\
& &
                                           + 126 x^3 ) \zeta(2)) H(1;x)
                     - \frac{3 (-14 \zeta(2) + 6 x \zeta(2) - 2 x^2 \zeta(2) + x^3 \zeta(2))}{8 (x-1)^4} H(2;x)  \nn\\
& &
                 + \frac{3 x \zeta(2)}{(x-1)^4} H(0, -1;x)
                 - \frac{1}{24 (x-1)^4 x} (11 + 22 x - 171 x^2 + 79 x^3 + 59 x^4 - ( 36 x  \nn\\
& &
                                                    + 270 x^2 + 216 x^3 ) \zeta(2)) H(0, 1;x)
                 - \frac{20 + 40 x - 66 x^2 - 3 x^3}{8 (x-1)^4} H(0, 0, 1;x)  \nn\\
& &
                 - \frac{1 + 5 x + 5 x^2 + x^3}{2 (x-1)^3 x} H(-1, 0, 1;x)
                 - \frac{13 + 82 x + 58 x^2}{24 (x-1)^2 x} H(1, 1;x)  \nn\\
& &
                 - \frac{22 + 56 x - 62 x^2 - 7 x^3}{8 (x-1)^4} H(0, 1, 1;x)
                 - \frac{10 + 23 x + 9 x^2}{8 (x-1)^3} H(1, 0, 1;x)  \nn\\
& &
                 + \frac{14 - 6 x + 2 x^2 - x^3}{8 (x-1)^4} H(2, 1, 1;x)
                 - \frac{2 + 31 x + 12 x^2}{4 (x-1)^4} H(0, 0, 0, 1;x)  \nn\\
& &
                 + \frac{6 x}{(x-1)^4} H(0, -1, 0, 1;x)
                 - \frac{2 + 7 x + 12 x^2}{4 (x-1)^4} ( 2 H(0, 0, 1, 1;x) - H(0, 1, 0, 1;x) )
                         \biggr]  \nn\\
& &
           + T_R N_l \biggl[ \frac{(25 + 6 L_{\mu})}{36 x} H(1;x)
                            + \frac{1}{6 x} ( H(0, 1;x) + 2 H(1, 1;x) )
                         \biggr]  \nn\\
& &
           + T_R N_h \biggl[ \frac{(54 - 91 x + 20 x^2 + 17 x^3 - ( 36 + 12 x - 52 x^2 + 4 x^3 ) \zeta(2)
                                   + 36 x \zeta(3))}{6 (x-1)^4} \nn\\
& &
                        + \frac{25 \! + \! 322 x \! +\!  25 x^2 \! + \! ( 6 \! - \! 12 x \! + \! 6 x^2) L_{\mu} }{36 (x-1)^2 x} H(1;x)
                    - \frac{1 \! + \! 21 x \! + \! 21 x^2 \! + \! x^3}{6 (x-1)^3 x} H(0, 1;x)  \nn\\
& &
                    - \frac{6 x}{(x-1)^4} H(0, 0, 1;x)
                         \biggr]
\, .
\label{G32l}
\eea

We checked our results for the form factors $G_i^{(2l)}(x)$ $i=1,2,3$ against 
the calculation of Martin Beneke, Tobias Huber, and Xin-Quing Li 
\cite{Beneke}  and we found complete analytical agreement.

\section{Ward Identities \label{Ward}}

We explicitly checked that the UV renormalized form factors satisfy the on-shell Ward
identity\footnote{It can be proved that the Ward identity is fulfilled already at the
level of master integrals, irrespectively on the analytic expression of the MIs
themselves.}
\vspace*{8mm}
\be
i q_\mu
\hspace*{2cm}
\hbox{
  \begin{picture}(0,0)(0,0)
\SetScale{1.}
  \SetWidth{.2}
  \Photon(-40,0)(0,0){3}{11}
\ArrowLine(10,10)(32,32)
\Text(-35,4)[cb]{$W^+,\mu$}
\Text(12,-33)[cb]{$b$}
\Text(12,26)[cb]{$u$}
  \SetWidth{1.5}
\ArrowLine(32,-32)(10,-10)
  \SetWidth{.9}
  \GCirc(0,0){15}{0.8}
\end{picture}}
\hspace*{.8cm}
- M_W
\hspace*{1.4cm}
\hbox{
  \begin{picture}(0,0)(0,0)
\SetScale{1.}
  \SetWidth{.2}
  \DashLine(-40,0)(0,0){3}{11}
\ArrowLine(10,10)(32,32)
\Text(-35,4)[cb]{$\phi^+$}
\Text(12,-33)[cb]{$b$}
\Text(12,26)[cb]{$u$}
  \SetWidth{1.5}
\ArrowLine(32,-32)(10,-10)
  \SetWidth{.9}
  \GCirc(0,0){15}{0.8}
\end{picture}}
\hspace*{1.2cm} =
  \SetWidth{.2}
%
0 \, , \label{WTI}
\ee
\vspace*{8mm}\\
where $\phi$ is the charged pseudo-Goldstone boson and the gray
circles represents the sum of all two-loop
one-particle-irreducible QCD corrections to the vertices. The
Lorentz index associated to the $W$-boson is saturated by the
boson momentum $q^\mu$.
 In order
to satisfy the relation in Eq.~(\ref{WTI}) it is necessary to
renormalize also the factor $m_b$ appearing in the tree-level
$\phi^+ u b $ coupling. The relevant NNLO mass counter term can be
found in \cite{vanMe}.

The two-loop corrections to the scalar coupling of the pseudo-Goldstone boson to
quark can be absorbed in a single form-factor S, defined as follows
\vspace*{5mm}\\
\be
\hbox{
  \begin{picture}(0,0)(0,0)
\SetScale{1.}
  \SetWidth{.2}
  \DashLine(-40,0)(0,0){3}{11}
\ArrowLine(10,10)(32,32)
\Text(-35,4)[cb]{$\phi^+$}
\Text(12,-33)[cb]{$b$}
\Text(12,26)[cb]{$u$}
  \SetWidth{1.5}
\ArrowLine(32,-32)(10,-10)
  \SetWidth{.9}
  \GCirc(0,0){15}{0.8}
\end{picture}}
\hspace*{1.2cm} = -
\frac{m_b}{M_W} \ S (q^2) \,
\overline{u}(p) \left( 1 - \gamma_5\right) u (P) \, .
\ee \\
\vspace*{2mm}\\
The UV renormalized form factor has the following perturbative expansion in
$\alpha_s$:
\be
S =\frac{i g_w}{2 \sqrt{2}} V_{ub}\left[S^{(0 l)} + \left(\frac{\alpha_s}{\pi}
\right)S^{(1 l)} + \left(\frac{\alpha_s}{\pi} \right)^2   S^{(2 l)} +
{\mathcal O}\left( \frac{\alpha_s^3}{\pi^3}\right) \right]\, , \label{Sexp}
\ee
with $S^{(0 l)} =1$.

The one-loop form factor $S^{(1 l)}$ is given by
\bea
S^{(1 l)} &=& C(d) \left( \frac{\mu^2}{m^2}\right)^{(4-d)/2}
C_F \sum_{i = -2}^{1} S^{(1 l,i)} (d-4)^i  + {\mathcal O}\left((d-4)^2
\right)\, .
\eea
After UV renormalization (including the renormalization of the Yukawa
$\phi^+ u b$ coupling), the coefficients of the expansion in $(d-4)$ are
\bea
S^{(1 l,-2)} &=& -1 \, , \\
S^{(1 l,-1)} &=&  \frac{5}{4} + H(1;x)\, , \\
S^{(1 l,0)} &=& - 1
          - \frac{1}{2 x} H(1 ;x)
          - H(1,1 ;x)
          - \frac{1}{2} H(0,1 ;x)\, , \\
S^{(1 l,1)} &=& 1
          + \frac{x+1}{4x} H(1 ;x)
          + \frac{1}{2 x} H(1,1 ;x)
          + \frac{1}{4 x} H(0,1 ;x)
          + H(1,1,1 ;x) \nn \\
& &
          + \frac{1}{2} H(1,0,1 ;x)
          + \frac{1}{2} H(0,1,1 ;x)
          + \frac{1}{4} H(0,0,1 ;x)  .
\eea

 The two-loop form factor $S^{(2 l)}$ is given by
\bea
S^{(2 l)} &=& C^2(d) \left( \frac{\mu^2}{m^2}\right)^{4-d}
C_F \sum_{i = -4}^{0} S^{(2 l,i)} (d-4)^i  + {\mathcal O}\left(d-4
\right)\, .
\eea
where the coefficient of the expansion in $(d-4)$ are
\bea
S^{(2 l,-4)} &=& C_F \frac{1}{2} \, ,  \\
S^{(2 l,-3)} &=& C_F \biggl[ - \frac{5}{4} - H(1;x)
                     \biggr]
               - C_A  \frac{11}{8}
               + T_R N_l  \frac{1}{2}
 \, , \\
S^{(2 l,-2)} &=& C_F \biggl[ \frac{57}{32} 
                             + \frac{2 + 5 x}{(4 x)} H(1;x) 
                             + \frac{1}{2} H(0, 1;x) 
			     + 2 H(1, 1;x)
                     \biggr]
               + C_A \biggl[\frac{49 + 9 \zeta(2)}{72}  \nn\\ 
& &
	                     - \frac{11}{12} L_{\mu} 
			     + \frac{11}{12} H(1;x)
                     \biggr]
               + T_R N_l \biggl[ -\frac{5}{18} 
	                          + \frac{1}{3} L_{\mu}  
				  - \frac{1}{3} H(1;x)
                     \biggr]
               + T_R N_h \biggl[ \frac{1}{3} L_{\mu}
                     \biggr]
\, , \\
S^{(2 l,-1)} &=& C_F \biggl[- \frac{3(47 + 8 \zeta(2) - 16 \zeta(3))}{64}  
                            - \frac{7 + 10 x}{8 x} H(1;x)
			    - \frac{2 + 5 x}{8 x} H(0, 1;x) \nn\\ 
& &
			    - \frac{6 \! + \! 5 x}{4 x} H(1, 1;x)\! 
			    - \frac{1}{4} H(0, 0, 1;x) \! 
			    - \frac{3}{2} H(0, 1, 1;x) \! 
			    - H(1, 0, 1;x) \! 
			    - 4 H(1, 1, 1;x)
                     \biggr]  \nn\\ 
& &
               + C_A \biggl[ \frac{(1549 \! + \! 1980 L_{\mu} \! \! - \! 396 L_{\mu}^2 \! \! + \! 972 \zeta(2) \! - \! 1188 \zeta(3))}{1728}
	                   \!  + \! \frac{67 \! + \! 66 L_{\mu} \! \! - \! 18 \zeta(2)}{72}  H(1;x)
                     \biggr] \nn\\ 
& &
               + T_R N_l \biggl[ \frac{-125 - 180 L_{\mu} + 36 L_{\mu}^2 - 108 \zeta(2)}{432}  
	                        - \frac{5 + 6 L_{\mu}}{18}  H(1;x)
                     \biggr] \nn\\ 
& &
               + T_R N_h \biggl[ \frac{- 5 L_{\mu} + L_{\mu}^2 - \zeta(2)}{12}  
	                        - \frac{1}{3} L_{\mu} H(1;x)
                     \biggr]
\, ,\nn \\
S^{(2 l,0)} &=& C_F \biggl[  \frac{831}{256} + \biggl(  \frac{3 (13 - 7 x)}{16 (x-1)}  + 
					\frac{3 \ln{(2)} (x-4) (7 x-8)}{8 (x-1)^2} \biggr) \zeta(2) 
					+ \biggl( \frac{9 (5 - 34 x + 11 x^2)}{80 (x-1)^2}  \nn\\ 
& &
                                        - \frac{4 K}{5} \biggr) \zeta^2(2) 
					- \frac{(74 - 96 x + 13 x^2) \zeta(3)}{32 (x-1)^2} 
	                   + \frac{(1 + 2 x + x^2) \zeta(2)}{2 (x-1) x} H(-1;x)  \nn\\ 
& &
			   - \frac{17 + 8 x - 25 x^2 - ( 12 - 78 x + 30 x^2 ) \zeta(2) 
			                  - ( 8 x - 8 x^2) \zeta(3)}{16 (x-1) x} H(1;x) \nn\\ 
& & 
			   - \frac{9 (4 - 4 x + x^2) \zeta(2)}{8 (x-1)^2} H(2;x) 
			   - \frac{\zeta(2)}{2}  H(0, -1;x) 
			   + \frac{1}{16 (x-1)^2 x} (11 + 12 x - 15 x^2  \nn\\ 
& & 
                           - 8 x^3 + 12 x \zeta(2) + 24 x^2 \zeta(2)) H(0, 1;x)
			   + \frac{4 - x + 5 x^2}{8 x^2} H(1, 1;x) \nn\\ 
& & 
			   + \frac{1 + 2 x + x^2}{(x-1) x} H(-1, 0, 1;x)
			   - \frac{2 - 13 x + 20 x^2 - 3 x^3}{16 (x-1)^2 x} H(0, 0, 1;x) \nn\\ 
& & 
			   + \frac{14 + 5 x}{4 x} H(1, 1, 1;x)
			   + \frac{10 - 33 x + 20 x^2 + 6 x^3}{8 (x-1)^2 x} H(0, 1, 1;x) \nn\\ 
& & 
			   - \frac{2 + 5 x - 4 x^2}{4 (x-1) x} H(1, 0, 1;x)
			   - \frac{3 (4 - 4 x + x^2)}{8 (x-1)^2} H(2, 1, 1;x)
			   - H(0, -1, 0, 1;x)  \nn\\ 
& & 
			   - \frac{5 - 2 x + 3 x^2}{8 (x-1)^2} H(0, 0, 0, 1;x) 
			   + \frac{1 - 10 x + 3 x^2}{4 (x-1)^2}  H(0, 0, 1, 1;x)
			   + \frac{7}{2} H(0, 1, 1, 1;x)  \nn\\ 
& & 
			   + \frac{4 - 4 x + 3 x^2}{4 (x-1)^2} H(0, 1, 0, 1;x)
			   - \frac{1}{2} H(1, 0, 0, 1;x) 
			   + 3 H(1, 0, 1, 1;x)  \nn\\ 
& & 
			   + 2 H(1, 1, 0, 1;x) 
			   + 8 H(1, 1, 1, 1;x)
                     \biggr]
               + C_A \biggl[    - \frac{54589}{20736} 
	                	- \frac{11 L_{\mu}}{12} 
				+ \frac{55 L_{\mu}^2}{192} 
				- \frac{11 L_{\mu}^3}{288}   \nn\\ 
& & 
				+ \biggl(   \frac{11 L_{\mu}}{96}  
			            - \frac{3 \ln{(2)} (x-4) (7 x-8)}{16 (x-1)^2}  
				    - \frac{1067 + 49 x}{576 (x-1)} \biggr) \zeta(2) 
				- \biggl( 
			             \frac{179 - 250 x + 125 x^2}{160 (x-1)^2}  \nn\\ 
& & 
                                         - \frac{2 K}{5} \biggr) \zeta^2(2) 
				+ \frac{(896 - 1324 x + 347 x^2) \zeta(3)}{576 (x-1)^2}
	                    - \frac{(1 + 2 x + x^2) \zeta(2)}{4 (x-1) x} H(-1;x) \nn\\ 
& & 
			    + \frac{1}{432 (x-1) x} (708 + 198 L_{\mu} - 466 x - 198 L_{\mu} x - 99 L_{\mu}^2 x 
			                           - 242 x^2 + 99 L_{\mu}^2 x^2  \nn\\ 
& & 
                                                   - 216 \zeta(2) + 738 x \zeta(2) 
						   - 684 x^2 \zeta(2) - 378 x \zeta(3) + 378 x^2 \zeta(3) ) H(1;x)  \nn\\ 
& & 
			    + \frac{9 (4 - 4 x + x^2) \zeta(2)}{(16 (x-1)^2)} H(2;x)
			    + \frac{1}{4} \zeta(2) H(0, -1;x) 
			    - \frac{1}{144 (x-1)^2 x} (66 + 56 x  \nn\\ 
& & 
                                                       + \! 66 L_{\mu} x \! - \! 211 x^2 \! - \! 132 L_{\mu} x^2 
			                           \! + \! 89 x^3 \! + \! 66 L_{\mu} x^3 \! - \! ( 126 x \! - \! 144 x^2  
						   \! + \! 72 x^3 )\zeta(2) ) H(0, 1;x)  \nn\\ 
& & 
			    - \frac{78 + 223 x + 132 L_{\mu} x - 72 x \zeta(2)}{144 x} H(1, 1;x)
			    - \frac{1 + 2 x + x^2}{2 (x-1) x} H(-1, 0, 1;x) \nn\\ 
& & 
			    - \frac{40 - 56 x + 7 x^2}{48 (x-1)^2} H(0, 0, 1;x)
			    - \frac{11}{6} H(1, 1, 1;x) 
			    - \frac{44 - 88 x + 53 x^2}{48 (x-1)^2} H(0, 1, 1;x) \nn\\ 
& & 
			    + \frac{29 - 35 x}{48 (x-1)} H(1, 0, 1;x)
			    + \frac{3 (4 - 4 x + x^2)}{16 (x-1)^2} H(2, 1, 1;x)
			    + \frac{1}{2} H(0, -1, 0, 1;x)  \nn\\ 
& & 
			    - \frac{1}{8 (x-1)^2} H(0, 0, 0, 1;x) 
			    - \frac{1}{4 (x-1)^2} H(0, 0, 1, 1;x)
			    + \frac{1}{8 (x-1)^2} H(0, 1, 0, 1;x)  \nn\\ 
& & 
			    + \frac{1}{2} H(1, 0, 0, 1;x)
                     \biggr] 
               + T_R N_l \biggl[ 
	                         \frac{1}{5184} (3893 + 1728 L_{\mu} - 540 L_{\mu}^2 + 72 L_{\mu}^3 + 3420 \zeta(2)  \nn\\ 
& & 
                                         - 216 L_{\mu} \zeta(2) 
	                                 + 720 \zeta(3))
	                        + \frac{48 + 18 L_{\mu} + 28 x - 9 L_{\mu}^2 x + 45 x \zeta(2)}{108 x} H(1;x) \nn\\ 
& & 
				+ \frac{3 + 5 x + 3 L_{\mu} x}{18 x} H(0, 1;x)
				+ \frac{3 + 5 x + 3 L_{\mu} x}{9 x} H(1, 1;x)
				+ \frac{1}{6} H(0, 0, 1;x)  \nn\\ 
& & 
				+ \frac{1}{3} H(0, 1, 1;x) 
				+ \frac{1}{3} H(1, 0, 1;x) 
				+ \frac{2}{3} H(1, 1, 1;x)
                     \biggr] 
               + T_R N_h \biggl[ \frac{L_{\mu}}{3} 
	                        - \frac{5 L_{\mu}^2}{48} 
				+ \frac{L_{\mu}^3}{72}   \nn\\ 
& & 
				+ \frac{11407 - 17630 x + 8527 x^2}{2592 (x-1)^2} 
				- \biggl( \frac{L_{\mu}}{24} 
				    - \frac{409 - 747 x + 651 x^2 - 185 x^3}{144 (x-1)^3} \biggr) \zeta(2)  \nn\\ 
& & 
				- \frac{7 \zeta(3)}{36}
	                        + \frac{1}{108 (x-1)^2 x} ( 48 + 18 L_{\mu} + 104 x - 36 L_{\mu} x 
				                      - 9 L_{\mu}^2 x - 112 x^2 + 18 L_{\mu} x^2  \nn\\ 
& & 
                                                      + 18 L_{\mu}^2 x^2 
						      + 56 x^3 - 9 L_{\mu}^2 x^3 + 9 x \zeta(2) - 18 x^2 \zeta(2) 
						      + 9 x^3 \zeta(2) ) H(1;x) \nn\\ 
& & 
				 - \frac{1}{18 (x-1)^3 x} (3 \! + \! 14 x \! + \! 3 L_{\mu} x \! - \! 9 L_{\mu} x^2 
				                      - 6 x^3 + 9 L_{\mu} x^3 
				                      + 5 x^4 - 3 L_{\mu} x^4) H(0, 1;x)\nn\\ 
& & 
			         + \frac{L_{\mu}}{3}  H(1, 1;x) 
				 + \frac{1}{6} H(0, 0, 1;x)
                     \biggr]  .
\eea

When written in terms of form factors, the Ward identity in Eq.~(\ref{WTI})
reads as follows:
\be
 2 G_1^{(2l)}(x)  + x  G_2^{(2l)}(x)  + G_3^{(2l)}(x) - 2 S^{(2l)}(x) = 0 \, .
\label{WTIFF}
\ee
It can be checked that the form factors presented in this paper fulfill
Eq.~(\ref{WTIFF}).

\section{Conclusions \label{concl}}

In this paper, we presented analytic expressions for the two-loop
QCD corrections to the decay process $b \to u \, W^{*}  \to u \, l \,
\bar{\nu}$. This process is important for the precise
determination of the CKM matrix element $V_{ub}$ and, therefore,
for the study of flavor  and CP violation within and beyond the
Standard Model of fundamental interactions.

The Lorentz structure of the process is parametrized in terms of three form
factors,  whose analytic expression are given in the form of a Laurent series of
$(d-4)$, where $d$ is the  space-time dimension. The coefficients of the series
are expressed in the well known functional basis of harmonic polylogarithms of a
single dimensionless variable. The result can be used in a SCET framework,
after combining it with the jet and soft functions already known in the literature, for a
phenomenological determination of $|V_{ub}|$. The results presented here are  the first
step towards a complete determination of the NNLO QCD corrections to the
heavy-to-light quark transition.

\section*{Acknowledgments}

We are grateful to U.~Aglietti, for proposing the subject of the paper, and to
M.~Beneke for allowing us to compare our results with the ones  obtained by his
group.     We thank U.~Aglietti and  P.~Gambino for carefully reading the
manuscript and providing us with valuable feedback. We wish to thank T.~Becher,
C.~Greub, T.~Gehrmann, and B.~Pecjak for useful  discussions. We are indebted
with R.~Boughezal and G.~Bell for several numerical and analytical  cross-checks
of the master integrals.
We are grateful to J.~Vermaseren  for his kind assistance in the use of {\tt
FORM} \cite{FORM}, and to the authors of the packages {\tt AIR} \cite{AIR} and
{\tt FIESTA} \cite{FIESTA}, that were employed in partial checks of the
calculation.

R.B. wishes to thank the Theoretical Physics Department of the University of 
Florence for kind hospitality during a part of this work.

This work  was supported  by the Swiss National Science Foundation (SNF) under
contract 200020-117602. \\

{\bf Note Added}: While this paper was with the editors, the results were independently
confirmed by three different groups \cite{Beneke,Greub,Bell}.


\appendix


\section{The Master Integrals \label{app1}}

\begin{figure}
\vspace*{10mm}
\bc
\[
\vcenter{
\hbox{
  \begin{picture}(0,0)(0,0)
\SetScale{1.}
  \SetWidth{.2}
\DashLine(-50,30)(-35,15){2}
\Line(-35,15)(20,15)
\Line(20,15)(35,15)
\Line(-35,-5)(20,15)
\Line(-35,-25)(-5,15)
\Text(0,-45)[c]{(a)}
  \SetWidth{1.5}
\Line(-35,-38)(-35,-23)
\Line(-35,-23)(-35,15)
\end{picture}}
}
\hspace{4.cm}
\vcenter{
\hbox{
  \begin{picture}(0,0)(0,0)
\SetScale{1.}
  \SetWidth{.2}
\DashLine(-50,30)(-35,15){2}
\Line(-35,15)(20,15)
\Line(20,15)(35,15)
\Line(-35,-5)(15,15)
\Line(-35,-25)(15,15)
\Text(0,-45)[c]{(b)}
  \SetWidth{1.5}
\Line(-35,-38)(-35,-23)
\Line(-35,-23)(-35,15)
\end{picture}}
}
\hspace{4.cm}
\vcenter{
\hbox{
  \begin{picture}(0,0)(0,0)
\SetScale{1.}
  \SetWidth{.2}
\DashLine(-50,30)(-35,15){2}
\Line(-35,15)(20,15)
\Line(20,15)(35,15)
\Line(-35,-5)(15,15)
\Line(-35,-25)(15,15)
\GCirc(-35,5){3}{.1}
\Text(0,-45)[c]{(c)}
  \SetWidth{1.5}
\Line(-35,-38)(-35,-23)
\Line(-35,-23)(-35,15)
\end{picture}}
}
\hspace{4.cm}
\vcenter{
\hbox{
  \begin{picture}(0,0)(0,0)
\SetScale{1.}
  \SetWidth{.2}
\DashLine(-50,30)(-35,15){2}
\Line(14.5,15)(35,15)
\CArc(14,-35)(50,90,167)
\Line(-35,-24.5)(-35,15)
\Text(0,-45)[c]{(d)}
  \SetWidth{1.5}
\CArc(-35,25)(50,270,348.5)
\Line(-35,-38)(-35,-24.5)
\Line(-35,15)(14.5,15)
\end{picture}}
}\]
%
%
%
\vspace{2.2cm}
\[\vcenter{
\hbox{
  \begin{picture}(0,0)(0,0)
\SetScale{1.}
  \SetWidth{.2}
\DashLine(-50,30)(-35,15){2}
\Line(14.5,15)(35,15)
\CArc(14,-35)(50,90,167)
\Line(-35,-24.5)(-35,15)
\GCirc(-16,15){3}{.1}
\Text(0,-45)[c]{(e)}
  \SetWidth{1.5}
\CArc(-35,25)(50,270,348.5)
\Line(-35,-38)(-35,-24.5)
\Line(-35,15)(14.5,15)
\end{picture}}
}
%
\hspace{4.cm}
%
\vcenter{
\hbox{
  \begin{picture}(0,0)(0,0)
\SetScale{1.}
  \SetWidth{.2}
\DashLine(-50,30)(-35,15){2}
\Line(14.5,15)(35,15)
\CArc(14,-35)(50,90,167)
\Line(-35,-24.5)(-35,15)
\GCirc(-19,2.5){3}{.1}
\Text(0,-45)[c]{(f)}
  \SetWidth{1.5}
\CArc(-35,25)(50,270,348.5)
\Line(-35,-38)(-35,-24.5)
\Line(-35,15)(14.5,15)
\end{picture}}
}
\hspace{4.cm}
%
\vcenter{
\hbox{
  \begin{picture}(0,0)(0,0)
\SetScale{1.}
  \SetWidth{.2}
\DashLine(-50,30)(-35,15){2}
\Line(14.5,15)(35,15)
\Line(-35,-24.5)(-35,15)
\Line(-35,15)(14.5,15)
\Text(0,-45)[c]{(g)}
  \SetWidth{1.5}
\CArc(-35,25)(50,270,348.5)
\Line(-35,-38)(-35,-24.5)
\CArc(-10.25,32)(30,214.5,325.5)
\end{picture}}
}
\hspace{4.cm}
\vcenter{
\hbox{
  \begin{picture}(0,0)(0,0)
\SetScale{1.}
  \SetWidth{.2}
\DashLine(-50,30)(-35,15){2}
\Line(14.5,15)(35,15)
\Line(-35,-24.5)(-35,15)
\Line(-35,15)(14.5,15)
\GCirc(-12,15){3}{.1}
\Text(0,-45)[c]{(h)}
  \SetWidth{1.5}
\CArc(-35,25)(50,270,348.5)
\Line(-35,-38)(-35,-24.5)
\CArc(-10.25,32)(30,214.5,325.5)
\end{picture}}
}\]
\vspace*{14mm}
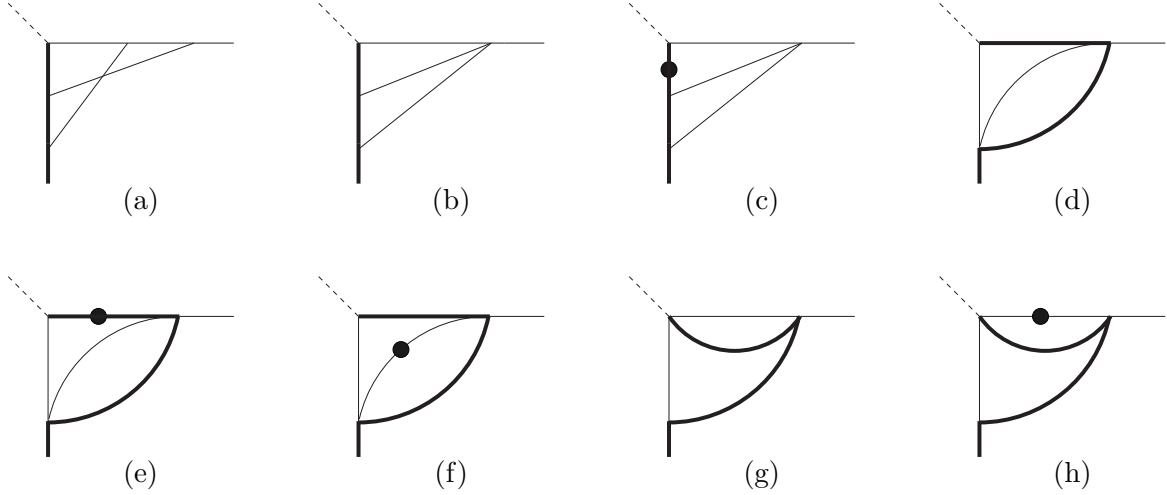
\caption{\label{fig3} Master Integrals needed for the Two-loop QCD corrections. Thick lines represent massive particles, thin lines represent massless ones.}
\ec
\efig
%

In this Appendix we collect the analytic expressions of the Master Integrals 
for the Feynman diagrams of Fig.~\ref{fig3}. We provide only eight of them, 
since the other MIs can be found in \cite{Bonciani:2008az,Fleischer:1999hp}.
It must be pointed out that the MIs (a)--(f) in Fig.~\ref{fig3} were already
calculated in \cite{Bell:2006tz}. We checked the analytic expressions that we obtained against the results in
\cite{Bell:2006tz} and we found
complete agreement.
Moreover, all the MIs were checked by comparing their numerical value to the results
obtained by direct numerical integration with the sector decomposition method. The
numerical integration was carried out by using the package {\tt FIESTA} (see \cite{FIESTA}).
The checks were done for several values of the variable $y$.

The explicit expression of the MIs depends on the chosen normalization of the
integration measure. The integration on the loop momenta is normalized as follows
\be
\int{\mathfrak
D}^dk = \frac{1}{C(d)} \left( \frac{\mu^2}{m^2}
\right)^{\frac{(d-4)}{2}} \int \frac{d^d k}{(4 \pi^2)^{\frac{(d-2)}{2}}} \, ,
\label{measure}
\ee
where $C(d)$ is defined in Eq.~(\ref{Cd}).
In Eq.~(\ref{measure}) $\mu$ stands
for the 't Hooft mass of dimensional regularization. The integration measure in
Eq.~(\ref{measure}) is chosen in such a way that the one-loop massive tadpole becomes
\be
\int{\mathfrak D}^dk \ \frac{1}{k^2+m^2} =
              \frac {m^2}{(d-2)(d-4)} \, .
\label{Tadpole}
\ee

In the expressions below, ${\mathcal K}$ is a rational number (its numerical value is ${\mathcal K}=3.32812 \pm 0.00002 \sim 213/64$),
while $a_4 = \mbox{Li}_4(1/2) = 0.51747906...$. $\zeta(2)$ and $\zeta(3)$ are the Riemann $\zeta$ function evaluated
in 2 and 3 respectively: $\zeta(2)=1.6449341...$, $\zeta(3)=1.2020569...$.

The expressions of the MIs are the following.

\begin{eqnarray}
\hbox{
  \begin{picture}(0,0)(0,0)
\SetScale{0.8}
  \SetWidth{.2}
\DashLine(-50,30)(-35,15){2}
\Line(-35,15)(20,15)
\Line(20,15)(35,15)
\Line(-35,-5)(20,15)
\Line(-35,-25)(-5,15)
  \SetWidth{1.5}
\Line(-35,-38)(-35,-23)
\Line(-35,-23)(-35,15)
\end{picture}}
& \hspace*{16mm} = & \frac{1}{m^4 (1+y)^2} \, \sum_{i=-4}^0 A_i \, (d-4)^i + {\mathcal O}(d-4) \, ,
\end{eqnarray}

\vspace*{5mm}

\bea
A_{-4} & = & \frac{1}{12}
\, , \\
A_{-3} & = & \frac{1}{6} H(-1;y)
\, , \\
A_{-2} & = &
          - \frac{7}{48} \zeta(2)
          + \frac{1}{3} H(-1,-1;y)
\, , \\
A_{-1} & = &
            \frac{89}{96} \zeta(3)
          - \frac{7}{24} \zeta(2) H(-1;y)
          + \frac{2}{3} H(-1,-1,-1;y)
\, , \\
A_{0} & = &
       - \frac{2}{5} \zeta^2(2) {\mathcal K}
          + \frac{65}{48} \zeta(3) H(-1;y)
          - \frac{7}{12} \zeta(2) H(-1,-1;y)
          + \frac{4}{3} H(-1,-1,-1,-1;y) \nn\\
&  &
          - \frac{1}{2} H(-1,0,0,-1;y)
\, .
\eea

\begin{eqnarray}
\hbox{
  \begin{picture}(0,0)(0,0)
\SetScale{0.8}
  \SetWidth{.2}
\DashLine(-50,30)(-35,15){2}
\Line(-35,15)(20,15)
\Line(20,15)(35,15)
\Line(-35,-5)(15,15)
\Line(-35,-25)(15,15)
  \SetWidth{1.5}
\Line(-35,-38)(-35,-23)
\Line(-35,-23)(-35,15)
\end{picture}}
& \hspace*{16mm} = & \frac{1}{m^2 (1+y)} \, B_0 + {\mathcal O}(d-4) \, ,
\end{eqnarray}

\vspace*{5mm}

\bea
B_0 & = &
            \frac{27}{160} \zeta^2(2)
          + \frac{3}{16} \zeta(2) H(0,-1;y)
          - \frac{1}{16} H(0,-1,0,-1;y)
          + \frac{1}{8} H(0,0,-1,-1;y) \nn\\
& &
          - \frac{1}{16} H(0,0,0,-1;y)
\, .
\eea

\begin{eqnarray}
\hbox{
  \begin{picture}(0,0)(0,0)
\SetScale{0.8}
  \SetWidth{.2}
\DashLine(-50,30)(-35,15){2}
\Line(-35,15)(20,15)
\Line(20,15)(35,15)
\Line(-35,-5)(15,15)
\Line(-35,-25)(15,15)
\GCirc(-35,5){3}{.1}
  \SetWidth{1.5}
\Line(-35,-38)(-35,-23)
\Line(-35,-23)(-35,15)
\end{picture}}
& \hspace*{16mm} = & \frac{1}{m^4 (1+y)} \, \sum_{i=-1}^1 C_i \, (d-4)^i + {\mathcal O}(d-4)^2 \, ,
\end{eqnarray}

\vspace*{5mm}

\bea
C_{-1} & = & \frac{1}{32} \zeta(2)
\, , \\
C_{0} & = &
            \frac{1}{64} \zeta(3)
          + \frac{1}{16} \zeta(2) H(-1;y)
          + \frac{1}{16} H(0,-1,-1;y)
          - \frac{1}{16} H(0,0,-1;y)
\, , \\
C_{1} & = &
          - \frac{9}{80} \zeta^2(2)
          + \frac{1}{32} \zeta(3) H(-1;y)
          - \frac{3}{32} \zeta(2) H(0,-1;y)
          + \frac{1}{8} \zeta(2) H(-1,-1;y) \nn\\
& &
          + \frac{1}{8} H(-1,0,-1,-1;y)
          - \frac{1}{8} H(-1,0,0,-1;y)
          + \frac{3}{16} H(0,-1,-1,-1;y) \nn\\
& &
          - \frac{1}{16} H(0,-1,0,-1;y)
          - \frac{9}{32} H(0,0,-1,-1;y)
          + \frac{5}{32} H(0,0,0,-1;y)
\, .
\eea

\begin{eqnarray}
\hbox{
  \begin{picture}(0,0)(0,0)
\SetScale{0.8}
  \SetWidth{.2}
\DashLine(-50,30)(-35,15){2}
\Line(14.5,15)(35,15)
\CArc(14,-35)(50,90,167)
\Line(-35,-24.5)(-35,15)
  \SetWidth{1.5}
\CArc(-35,25)(50,270,348.5)
\Line(-35,-38)(-35,-24.5)
\Line(-35,15)(14.5,15)
\end{picture}}
& \hspace*{16mm} = & \sum_{i=-2}^1 D_i \, (d-4)^i + {\mathcal O}(d-4)^2 \, ,
\end{eqnarray}

\vspace*{5mm}

\bea
D_{-2} & = & \frac{1}{8}
\, , \\
D_{-1} & = &
          - \frac{5}{16}
          + \frac{(1+y)}{8y} H(-1;y)
\, , \\
D_{0} & = &
            \frac{19}{32}
          + \frac{1}{16} \zeta(2)
          - \frac{5}{16} H(-1;y)
          + \frac{3}{16} H(-1,-1;y)
          - \frac{1}{16} H(0,-1;y)\nn\\
& &
     + \frac{1}{y} \Biggl[
          - \frac{5}{16} H(-1;y)
          + \frac{3}{16} H(-1,-1;y)
          - \frac{1}{8} H(0,-1;y)
               \Biggr]
     + \frac{1}{(1+y)} \Biggl[
            \frac{5}{64} \zeta(3) \nn\\
& &
          - \frac{3}{16} \zeta(2) \ln(2)
          - \frac{3}{16} \zeta(2) H(-2;y)
          - \frac{1}{16} H(-2,-1,-1;y)
          - \frac{1}{16} H(0,-1,-1;y)\nn\\
& &
          + \frac{1}{16} H(0,0,-1;y)
               \Biggr]
\, , \\
D_{1} & = &
          - \frac{65}{64}
          - \frac{7}{128} \zeta(3)
          - \frac{5}{32} \zeta(2)
          - \frac{3}{32} \zeta(2) \ln(2)
          + \frac{19}{32} H(-1;y)
          + \frac{5}{32} \zeta(2) H(-1;y) \nn\\
& &
          - \frac{3}{32} \zeta(2) H(-2;y)
          + \frac{5}{32} H(0,-1;y)
          - \frac{15}{32} H(-1,-1;y)
          - \frac{1}{32} H(-2,-1,-1;y) \nn\\
& &
          + \frac{5}{16} H(-1,-1,-1;y)
          - \frac{1}{8} H(-1,0,-1;y)
          - \frac{3}{32} H(0,-1,-1;y)
          + \frac{1}{16} H(0,0,-1;y) \nn\\
& &
      + \frac{1}{y} \Biggl[
            \frac{19}{32} H(-1;y)
          + \frac{3}{32} \zeta(2) H(-1;y)
          - \frac{15}{32} H(-1,-1;y)
          + \frac{5}{16} H(0,-1;y) \nn\\
& &
          \!+ \frac{5}{16} H(-1,-1,-1;y)
          \!- \frac{3}{16} H(-1,0,-1;y)
          \!- \frac{3}{16} H(0,-1,-1;y)
          \!+ \frac{1}{8} H(0,0,-1;y)
               \Biggr] \nn\\
& &
      + \frac{1}{(1+y)} \Biggl[
          - \frac{1}{96} \ln^4(2)
          - \frac{5}{128} \zeta(3)
          + \frac{3}{32} \zeta(2) \ln(2)
          + \frac{1}{16} \zeta(2) \ln^2(2)
          + \frac{33}{640} \zeta^2(2)
          - \frac{1}{4} a_4 \nn\\
& &
          - \Bigl( \frac{5}{64} \zeta(3)
          - \frac{3}{16} \zeta(2) \ln(2) \Bigr) H(-1;y)
          + \Bigl( \frac{7}{32} \zeta(3)
          + \frac{3}{32} \zeta(2) \Bigr) H(-2;y)\nn\\
& &
          - \frac{3}{32} \zeta(2) H(0,-1;y)
          + \frac{3}{16} \zeta(2) H(-1,-2;y)
          - \frac{5}{32} \zeta(2) H(-2,-1;y)\nn\\
& &
          + \frac{1}{32} H(-2,-1,-1;y)
          + \frac{1}{32} H(0,-1,-1;y)
          - \frac{1}{32} H(0,0,-1;y)\nn\\
& &
          - \frac{3}{16} H(-2,-1,-1,-1;y)
          + \frac{1}{16} H(-2,-1,0,-1;y)
          + \frac{1}{16} H(-1,-2,-1,-1;y)\nn\\
& &
          + \frac{1}{16} H(-1,0,-1,-1;y)
          - \frac{1}{16} H(-1,0,0,-1;y)
          - \frac{3}{16} H(0,-1,-1,-1;y)\nn\\
& &
          + \frac{1}{8} H(0,-1,0,-1;y)
          + \frac{1}{8} H(0,0,-1,-1;y)
          - \frac{3}{32} H(0,0,0,-1;y)
               \Biggr]
\, .
\eea

\begin{eqnarray}
\hbox{
  \begin{picture}(0,0)(0,0)
\SetScale{0.8}
  \SetWidth{.2}
\DashLine(-50,30)(-35,15){2}
\Line(14.5,15)(35,15)
\CArc(14,-35)(50,90,167)
\Line(-35,-24.5)(-35,15)
\GCirc(-16,15){3}{.1}
  \SetWidth{1.5}
\CArc(-35,25)(50,270,348.5)
\Line(-35,-38)(-35,-24.5)
\Line(-35,15)(14.5,15)
\end{picture}}
& \hspace*{16mm} = & \frac{1}{m^2} \, \sum_{i=-1}^2 E_i \, (d-4)^i + {\mathcal O}(d-4)^3 \, ,
\end{eqnarray}

\vspace*{5mm}

\bea
E_{-1} & = &
          - \frac{1}{8y} H(-1;y)
\, , \\
E_{0} & = &
         \frac{1}{8y} \Biggl[
            H(-1;y)
          - \frac{3}{2} H(-1,-1;y)
          + H(0,-1;y)
               \Biggr]\nn\\
& &
        - \frac{1}{16(2+y)} \Biggl[
            3 \zeta(2)
          + H(-1,-1;y)
               \Biggr]
\, , \\
E_{1} & = &
         \frac{1}{y} \Biggl[
          - \frac{1}{8} H(-1;y)
          - \frac{3}{32} \zeta(2) H(-1;y)
          + \frac{3}{16} H(-1,-1;y) \nn\\
& &
          - \frac{5}{16} H(-1,-1,-1;y)
          + \frac{3}{16} H(-1,0,-1;y)
          - \frac{1}{8} H(0,-1;y)
          + \frac{3}{16} H(0,-1,-1;y)\nn\\
& &
          - \frac{1}{8} H(0,0,-1;y)
               \Biggr]
        + \frac{1}{(1+y)} \Biggl[
          - \frac{5}{64} \zeta(3)
          + \frac{3}{16} \zeta(2) \ln(2)
          + \frac{3}{16} \zeta(2) H(-2;y)\nn\\
& &
          + \frac{1}{16} H(-2,-1,-1;y)
          + \frac{1}{16} H(0,-1,-1;y)
          - \frac{1}{16} H(0,0,-1;y)
               \Biggr]
        + \frac{1}{(2+y)} \Biggl[
            \frac{7}{32} \zeta(3)\nn\\
& &
          + \frac{3}{16} \zeta(2)
          - \frac{5}{32} \zeta(2) H(-1;y)
          + \frac{1}{16} H(-1,-1;y)
          - \frac{3}{16} H(-1,-1,-1;y)\nn\\
& &
          + \frac{1}{16} H(-1,0,-1;y)
               \Biggr]
\, , \\
E_{2} & = &
        \frac{1}{y} \Biggl[
           \Bigl( \frac{1}{8}
          + \frac{9}{128} \zeta(3)
          + \frac{3}{32} \zeta(2)
          + \frac{3}{32} \zeta(2) \ln(2) \Bigr) H(-1;y)
          + \Bigl( \frac{1}{8}
          + \frac{3}{32} \zeta(2) \Bigr) H(0,-1;y) \nn\\
& &
          - \Bigl( \frac{3}{16}
          + \frac{7}{32} \zeta(2)\Bigr)  H(-1,-1;y)
          + \frac{3}{32} \zeta(2) H(-1,-2;y)
          - \frac{3}{16} H(-1,0,-1;y) \nn\\
& &
          + \frac{1}{8} H(0,0,-1;y)
          - \frac{3}{16} H(0,-1,-1;y)
          + \frac{5}{16} H(-1,-1,-1;y)\nn\\
& &
          + \frac{1}{32} H(-1,-2,-1,-1;y)
          - \frac{9}{16} H(-1,-1,-1,-1;y)
          + \frac{5}{16} H(-1,-1,0,-1;y)\nn\\
& &
          + \frac{5}{16} H(-1,0,-1,-1;y)
          - \frac{7}{32} H(-1,0,0,-1;y)
          + \frac{5}{16} H(0,-1,-1,-1;y) \nn\\
& &
          - \frac{3}{16} H(0,-1,0,-1;y)
          - \frac{3}{16} H(0,0,-1,-1;y)
          + \frac{1}{8} H(0,0,0,-1;y)
               \Biggr]\nn\\
& &
        + \frac{1}{(1+y)} \Biggl[
            \frac{1}{96} \ln^4(2)
          + \frac{5}{64} \zeta(3)
          - \frac{3}{16} \zeta(2) \ln(2)
          - \frac{1}{16} \zeta(2) \ln^2(2)
          - \frac{33}{640} \zeta^2(2)
          + \frac{1}{4} a_4\nn\\
& &
          + \Bigl( \frac{5}{64} \zeta(3)
          - \frac{3}{16} \zeta(2) \ln(2) \Bigr)  H(-1;y)
          - \Bigl( \frac{7}{32} \zeta(3)
          + \frac{3}{16} \zeta(2) \Bigr) H(-2;y)\nn\\
& &
          + \frac{3}{32} \zeta(2) H(0,-1;y)
          + \frac{5}{32} \zeta(2) H(-2,-1;y)
          - \frac{3}{16} \zeta(2) H(-1,-2;y)\nn\\
& &
          - \frac{1}{16} H(0,-1,-1;y)
          + \frac{1}{16} H(0,0,-1;y)
          - \frac{1}{16} H(-2,-1,-1;y) \nn\\
& &
          + \frac{3}{16} H(-2,-1,-1,-1;y)
          - \frac{1}{16} H(-2,-1,0,-1;y)
          - \frac{1}{16} H(-1,-2,-1,-1;y) \nn\\
& &
          - \frac{1}{16} H(-1,0,-1,-1;y)
          + \frac{1}{16} H(-1,0,0,-1;y)
          + \frac{3}{16} H(0,-1,-1,-1;y) \nn\\
& &
          - \frac{1}{8} H(0,-1,0,-1;y)
          - \frac{1}{8} H(0,0,-1,-1;y)
          + \frac{3}{32} H(0,0,0,-1;y)
               \Biggr] \nn\\
& &
        + \frac{1}{(2+y)} \Biggl[
          - \frac{7}{32} \zeta(3)
          - \frac{3}{16} \zeta(2)
          - \frac{45}{128} \zeta^2(2)
          + \Bigl( \frac{1}{8} \zeta(3)
          + \frac{5}{32} \zeta(2)\nn\\
& &
          + \frac{3}{16} \zeta(2) \ln(2) \Bigr) H(-1;y)
          - \Bigl( \frac{1}{16}
          + \frac{9}{32} \zeta(2) \Bigr) H(-1,-1;y)
          + \frac{3}{16} \zeta(2) H(-1,-2;y)\nn\\
& &
          + \frac{3}{16} H(-1,-1,-1;y)
          - \frac{1}{16} H(-1,0,-1;y)
          + \frac{1}{16} H(-1,-2,-1,-1;y) \nn\\
& &
          - \frac{7}{16} H(-1,-1,-1,-1;y)
          + \frac{3}{16} H(-1,-1,0,-1;y)
          + \frac{5}{32} H(-1,0,-1,-1;y) \nn\\
& &
          - \frac{1}{8} H(-1,0,0,-1;y)
               \Biggr]
\, .
\eea

\begin{eqnarray}
\hbox{
  \begin{picture}(0,0)(0,0)
\SetScale{0.8}
  \SetWidth{.2}
\DashLine(-50,30)(-35,15){2}
\Line(14.5,15)(35,15)
\CArc(14,-35)(50,90,167)
\Line(-35,-24.5)(-35,15)
\GCirc(-19,2.5){3}{.1}
  \SetWidth{1.5}
\CArc(-35,25)(50,270,348.5)
\Line(-35,-38)(-35,-24.5)
\Line(-35,15)(14.5,15)
\end{picture}}
& \hspace*{16mm} = & \frac{1}{m^2 (1+y)}\sum_{i=-1}^1 F_i \, (d-4)^i + {\mathcal O}(d-4)^2 \, ,
\end{eqnarray}

\vspace*{5mm}

\bea
F_{-1} & = &
            \frac{1}{8} \zeta(2)
          + \frac{1}{8} H(0,-1;y)
\, , \\
F_{0} & = &
          - \frac{7}{64} \zeta(3)
          - \frac{3}{16} \zeta(2) \ln(2)
          + \frac{1}{8} \zeta(2) H(-1;y)
          - \frac{3}{16} \zeta(2) H(-2;y)
          - \frac{1}{16} H(-2,-1,-1;y) \nn\\
& &
          + \frac{1}{8} H(-1,0,-1;y)
          + \frac{3}{16} H(0,-1,-1;y)
          - \frac{1}{8} H(0,0,-1;y)
\, , \\
F_{1} & = &
          - \frac{1}{96} \ln^4(2)
          + \frac{1}{16} \zeta(2) \ln^2(2)
          + \frac{227}{640} \zeta^2(2)
          - \frac{1}{4} a_4
          - \frac{3}{16} \zeta(3) H(-1;y)
          + \frac{7}{32} \zeta(3) H(-2;y) \nn\\
& &
          + \frac{\zeta(2)}{32} \bigl[
            3 H(0,-1;y)
          + 4 H(-1,-1;y)
          - 5 H(-2,-1;y) \bigr]
          - \frac{3}{16} H(-2,-1,-1,-1;y) \nn\\
& &
          + \frac{1}{16} H(-2,-1,0,-1;y)
          + \frac{1}{8} H(-1,-1,0,-1;y)
          + \frac{1}{4} H(-1,0,-1,-1;y) \nn\\
& &
          - \frac{3}{16} H(-1,0,0,-1;y)
          + \frac{5}{16} H(0,-1,-1,-1;y)
          - \frac{3}{16} H(0,-1,0,-1;y) \nn\\
& &
          - \frac{3}{16} H(0,0,-1,-1;y)
          + \frac{1}{8} H(0,0,0,-1;y)
\, .
\eea

\begin{eqnarray}
\hbox{
  \begin{picture}(0,0)(0,0)
\SetScale{0.8}
  \SetWidth{.2}
\DashLine(-50,30)(-35,15){2}
\Line(14.5,15)(35,15)
\Line(-35,-24.5)(-35,15)
\Line(-35,15)(14.5,15)
  \SetWidth{1.5}
\CArc(-35,25)(50,270,348.5)
\Line(-35,-38)(-35,-24.5)
\CArc(-10.25,32)(30,214.5,325.5)
\end{picture}}
& \hspace*{16mm} = & \sum_{i=-2}^1 G_i \, (d-4)^i + {\mathcal O}(d-4)^2 \, ,
\end{eqnarray}

\vspace*{5mm}

\bea
G_{-2} & = & \frac{1}{8}
\, , \\
G_{-1} & = & - \frac{5}{16}
\, , \\
G_{0} & = &
            \frac{19}{32}
          - \frac{1}{16} \zeta(2)
          - \frac{1}{16} H(0,-1;y)
          - \frac{1}{8(y+1)} \bigl[ z3 + H(0,0,-1;y) \bigr]
\, , \\
G_{1} & = &
          - \frac{65}{64}
          + \frac{3}{32} \zeta(3)
          + \frac{5}{32} \zeta(2)
          - \frac{1}{16} \zeta(2) H(-1;y)
          + \frac{5}{32} H(0,-1;y)
          - \frac{1}{16} H(-1,0,-1;y) \nn\\
& &
          - \frac{1}{8} H(0,-1,-1;y)
          + \frac{3}{32} H(0,0,-1;y)
        + \frac{1}{(1+y)} \Biggl[
            \frac{1}{16} \zeta(3)
          + \frac{7}{160} \zeta^2(2)\nn\\
& &
          + \frac{1}{8} \zeta(3) H(-1;y)
          - \frac{1}{16} \zeta(2) H(0,-1;y)
          + \frac{1}{16} H(0,0,-1;y)
          + \frac{1}{8} H(-1,0,0,-1;y)\nn\\
& &
          - \frac{1}{16} H(0,-1,0,-1;y)
          - \frac{1}{4} H(0,0,-1,-1;y)
                \Biggr]
\, .
\eea

\begin{eqnarray}
\hbox{
  \begin{picture}(0,0)(0,0)
\SetScale{0.8}
  \SetWidth{.2}
\DashLine(-50,30)(-35,15){2}
\Line(14.5,15)(35,15)
\Line(-35,-24.5)(-35,15)
\Line(-35,15)(14.5,15)
\GCirc(-12,15){3}{.1}
  \SetWidth{1.5}
\CArc(-35,25)(50,270,348.5)
\Line(-35,-38)(-35,-24.5)
\CArc(-10.25,32)(30,214.5,325.5)
\end{picture}}
& \hspace*{16mm} = & \frac{1}{m^2 (1+y)} \, \sum_{i=-1}^1 J_i \, (d-4)^i + {\mathcal O}(d-4)^2 \, ,
\end{eqnarray}

\vspace*{5mm}

\bea
J_{-1} & = &
            \frac{1}{8} \zeta(2)
          + \frac{1}{8} H(0,-1;y)
\, , \\
J_{0} & = &
          - \frac{3}{16} \zeta(3)
          + \frac{1}{8} \zeta(2) H(-1;y)
          + \frac{1}{8} H(-1,0,-1;y)
          + \frac{1}{4} H(0,-1,-1;y) \nn\\
& &
          - \frac{3}{16} H(0,0,-1;y)
\, , \\
J_{1} & = &
            \frac{23}{160} \zeta^2(2)
          - \frac{1}{16} \zeta(3) H(-1;y)
          + \frac{1}{8} \zeta(2) H(-1,-1;y)
          + \frac{1}{8} H(-1,-1,0,-1;y) \nn\\
& &
          + \frac{1}{4} H(-1,0,-1,-1;y)
          - \frac{1}{16} H(-1,0,0,-1;y)
          + \frac{1}{2} H(0,-1,-1,-1;y) \nn\\
& &
          - \frac{1}{4} H(0,-1,0,-1;y)
          - \frac{3}{8} H(0,0,-1,-1;y)
          + \frac{1}{32} H(0,0,0,-1;y)
\, .
\eea

\end{document}